\documentclass[12pt,preprint]{aastex}

\newcommand{\vect}[1]{\mbox{\boldmath$#1$}}
\newcommand{\dfrac}[2]{{\displaystyle \frac{#1}{#2}}  }

\newcommand{\eqref}[1]{(\ref{#1})}

\def\lesssim{\mathrel{\hbox{\rlap{\hbox{\lower4pt\hbox{$\sim$}}}\hbox{$<$}}}}
\def\gtrsim{\mathrel{\hbox{\rlap{\hbox{\lower4pt\hbox{$\sim$}}}\hbox{$>$}}}}

\slugcomment{Accepted for Publication in ApJ}

\shorttitle{Orbit in a Protoplanetary Disk}
\shortauthors{Muto and Inutsuka}

\begin{document}

\title{Orbital Evolution of a Particle
 Interacting with a Single Planet in a Protoplanetary Disk}

\author{Takayuki Muto\altaffilmark{1} and Shu-ichiro Inutsuka}
\affil{Department of Physics, Kyoto University, \\
Kitashirakawa-oiwake-cho, Sakyo-ku, Kyoto, 606-8502, Japan}
\email{muto@tap.scphys.kyoto-u.ac.jp}

\altaffiltext{1}{JSPS Research Fellow}

\begin{abstract}
We investigate the motion of a particle around a low mass planet
 embedded in a non-turbulent gaseous disk.  We take into account the
 effect of the gas
 structure that is modified by the gravitational
 interaction between the planet.  We derive an analytic formula that
 describes the change of the semi-major axis of the particle due to the
 encounter with the planet using 
 local approximation in distant encounter regime.  Our final formula
 includes the effects of steady, axisymmetric radial gas 
 flow, the global gas pressure gradient in the disk, planet gravity, and
 the structure of the gas flow
 modified by the planet's gravity.   We compare
 the analytic results with numerical calculations, and indicate
 that our formula well describes the secular evolution of the dust
 particles' semi-major axes well, especially
 for small particles with large drag coefficient.  We discuss the
 conditions for dust gap opening around a low mass planet and radial
 distribution of dust particles.  Our formula may provide a
 useful tool for calculating radial distribution of particles
 in a disk around the planet.
\end{abstract}

\keywords{planet and satellites: formation --- solar system: formation
--- celestial mechanics --- methods: analytical}

\section{Introduction} \label{intro}

The motion of small particles in a gas disk has been investigated in
detail since this is essential for planet formation theory.  
In a number of work, an axisymmetric gas disk is assumed 
and the effects of the structure of the gas disk on
particle orbit are investigated, neglecting the effects of planet's
gravitational force (Adachi et al. 1976, Weidenschilling 1977).  Their
results show that particles with 
stopping timescale of the order of Kepler timescale migrate towards the
central star rapidly, thereby imposing a serious barrier on the formation
of planetesimals.

Although this poses a serious question on the formation of
planetesimals, there is another aspect of this fact in the formation of
gas giant core. 
Gas giant core should reach the critical mass to accrete nearby
gas rapidly (Mizuno et al. 1978), and its formation timescale 
seems to be very long (Pollack et al. 1993). 
If the global pressure gradient exerted on the gas disk will
force the particles to migrate towards the central star, the particles
at outer disk may reach the orbit of an already formed core and may
accrete onto it.  Kary et al. (1993) have shown that this process seems
to be possible but most of the particles may miss the planet since they
migrate very rapidly.  Their discussion is based on the three-body
problem including the effect of axisymmetric gas disk.

The assumption of axisymmetric gas disk is a simplification in order to
make the calculation tractable.  The planet gravitationally interacts
with the gas disk and changes the structure of the disk around it,
producing the spiral density wave (see e.g., Goldreich and Tremaine
1979).  Only recently does appear the 
investigation of particle motion around 
the planet embedded in a gas disk, which fully takes into account
the structure around the planet (Paardekooper and Mellema 2004, 2006,
Paardekooper 2007, Fouchet et al. 2007, Lyra et al. 2008).
  For instance, Paardekooper and Mellema (2004, 2006)
and Paardekooper (2007) numerically
calculated the particle motion around a high mass planet in the
presence of a gas disk and discussed the accretion of the particles onto
the planet and the formation of gap of particles of different size.
They concluded that minimum mass of the planet that opens up a
dust gap was $0.05M_J$, thereby indicating the possibility of detecting
a low mass planet by future observations by observing a dust gap.

Many of such works involve numerical studies on particle motion.  
In this paper, we present an analytic investigation of the motion of
particles around a low mass planet embedded in a gas disk, taking into
account of the structure around the planet.
 This is the
 full analytic solution of a particle motion around the planet in which
 the structure of the gas disk is fully taken into account.
  We consider a vertically
averaged disk and perform two-dimensional analysis for simplicity.
  We use the results of
(quasi-)linear calculation of the gas structure around the planet and
calculate the distant encounter between a particle and a planet embedded
in a gaseous disk.  We derive a formula which describes the secular
change of the semi-major axis of the particle caused by the encounter
with the planet using Hill's equations.  
It is not possible to describe the resonant capture discussed
by Weidenschilling and Davis (1985) or Kary et al. (1993); however, we
take into account the structure of the gaseous disk that is modified by
the planet, which is not considered in most of the previous analytic
works.  Our 
paper is complementary with previous numerical studies in two respects. 
Firstly, our analysis is focused on low mass planet, and secondly, our
treatment is analytic.  It is challenging to follow numerically the
motion of dust particles embedded in a gas disk for long timescale
(e.g., $10^5$ Kepler time), solving gas and the motion of dust
simultaneously. 
Therefore, analytic treatment may be necessary in order to understand
the secular motion of dust particles.  Moreover, basic physical
processes in the problem become clear by analytic treatment.

%%%%%%%Modified in Revision%%%%%%%%
The plan of this paper is as follows.  In Section \ref{analytic}, we
show analytic treatment of the problem and derive the
formula for the evolution of semi-major axis, equation
\eqref{deltab_final}.  This formula includes the effects of global
pressure gradient of gas disk, steady, axisymmetric mass accretion flow
of the gas, gravitational force from the planet, and
the gas velocity structure modified by the planet.  We then compare our
results with numerical calculations in Sections \ref{numerical} and
\ref{global}, and the limitations of our analytic approach are
discussed. 
In Section \ref{dustgap}, we discuss the conditions for dust gap
formation and long-term behavior of dust particles whose orbits are
close to an embedded planet.  Section
\ref{summary} is for summary and future prospects.
%%%%%%%Modified in Revision end%%%%%%%%

We expect that our results, equation \eqref{deltab_final}, can be used
as a tool for estimate radial motion of the particle
in a disk around the planet, since this is
written in an analytic form.  It may be possible to calculate the radial
distribution of particles of various sizes using this formula and
derive, for example, the opacity of the disk.  We recommend readers
who need only the final results of the particle motion just to refer
equation \eqref{deltab_final} directly and go to Section
\ref{discuss:dustdistribution} where we demonstrate the calculation of
radial dust distribution around a planet.

%%%%%%%Added in Revision%%%%%%%
Before proceeding to the details of calculations, we note that there
are two relevant length scales in the problem at hand: pressure scale
height $H$ of the gas disk and Hill's radius of the planet, $r_{\rm H}$.  
Pressure scale height is an important quantity for the
structure of gas modified by the gravitational force of the planet,
while Hill's radius is important for the orbit of particles.  
  Pressure scale height that is given by sound speed of the gas $c$
  divided by Keplerian 
angular velocity and Hill's radius of the planet are related by  
\begin{equation}
 \dfrac{r_{\rm H}}{H} 
  \sim \dfrac{r_{\rm p}}{H}  
   \left( \dfrac{M_{\rm p}}{M_{\rm c}}  \right)^{1/3}
  = \left( \dfrac{GM_{\rm p}}{Hc^2} \right)^{1/3},
\end{equation}
where $r_{\rm p}$ is the semi-major axis of the planet, $M_{\rm p}$ is
the mass of the planet, and $M_{\rm c}$ is the mass of the central star.
For low mass planets 
considered in this paper, $r_{\rm H}/H$ is smaller than unity.
%%%%%%%Added in Revision end%%%%%%%

\section{Analytic Consideration} \label{analytic}

In this section, we analytically investigate the particle motion under
the influence of gravitational force by the protoplanet and gas drag
using local approximation.  We show that particles with friction time of
the order Kepler time will migrate towards the planet because of the gas
drag, although gravitational scattering tends to repel the particle from
the planet.  This mechanism is different from
the inward migration of dust particles caused by the velocity difference
between the gas and dust [see e.g., Adachi et al. (1976) or
Weidenschilling (1977)].  We also derive the effects of the modification
of gas velocity due to the gravitational force of the planet.

\subsection{Basic Equations}
\label{subsec:analytic_equation}

We use Hill's approximation to investigate the encounter between the
particle and the planet.  We consider the motion only in $(x,y)$ plane
where $x$ is in radial direction and $y$ is in the azimuthal direction.
We assume the planet is fixed at the origin $(x,y)=(0,0)$ in this
coordinate and neglect the mass of the particle.  We write the velocity
of the gas $\vect{v}_{\rm g}$ by
\begin{equation}
 \vect{v}_{\rm g} = -\dfrac{3}{2} \Omega_{\rm p} x \vect{e}_{y} 
  + \delta \vect{v}_{\rm g},
\end{equation}
where $\Omega_{\rm p}$ is the Keplerian angular velocity of the planet
and $\vect{e}_y$ is the unit vector in the $y$-direction. 
 The first term
represents the Keplerian motion around the central star and the second
term $\delta \vect{v}_{\rm g}$ 
represents the deviation from this motion, which,
for instance, may be caused by the global pressure gradient of the
disk.

The equations of motion of the particle are
\begin{equation}
 \ddot{x} - 2\Omega_{\rm p} \dot{y} = 3 \Omega_{\rm p}^2 x - \nu \dot{x}
  + F_x
  \label{EoM_x}
\end{equation}
\begin{equation}
 \ddot{y} + 2\Omega_{\rm p} \dot{x} = - \nu \left( \dot{y} + \dfrac{3}{2}
  \Omega_{\rm p} x \right) +  F_y
\label{EoM_y}
\end{equation}
where $\nu$ stands for the drag coefficient. 
%%%%%%Added in revision%%%%%%%%
 The quantity $\nu$ is a reciprocal of the stopping time of a particle.
 Note that 
 there are various notations for this parameter.  In Weidenschilling
 (1977), stopping time is denoted by $t_e$ so $\nu$ in this paper is
 equal to $1/t_e$.  In Adachi et al. (1976),
 stopping time corresponds to their $A$ times gas density times relative
 velocity. 
%%%%%%Added in revision end%%%%%%%%
We write the sum of
the force exerted by the planet and the part of the friction force
caused by the non-Keplerian motion of the gas by $\vect{F}$,
\begin{equation}
 \vect{F}(x,y) = \nabla \frac{GM_{\rm p}}{\sqrt{x^2 + y^2}}
  + \nu \delta \vect{v}_{\rm g} (x,y) ,
  \label{force_general}
\end{equation}
where $G$ is the gravitational constant and $M_{\rm p}$ denotes the mass
of the planet.

\begin{figure}
 \plotone{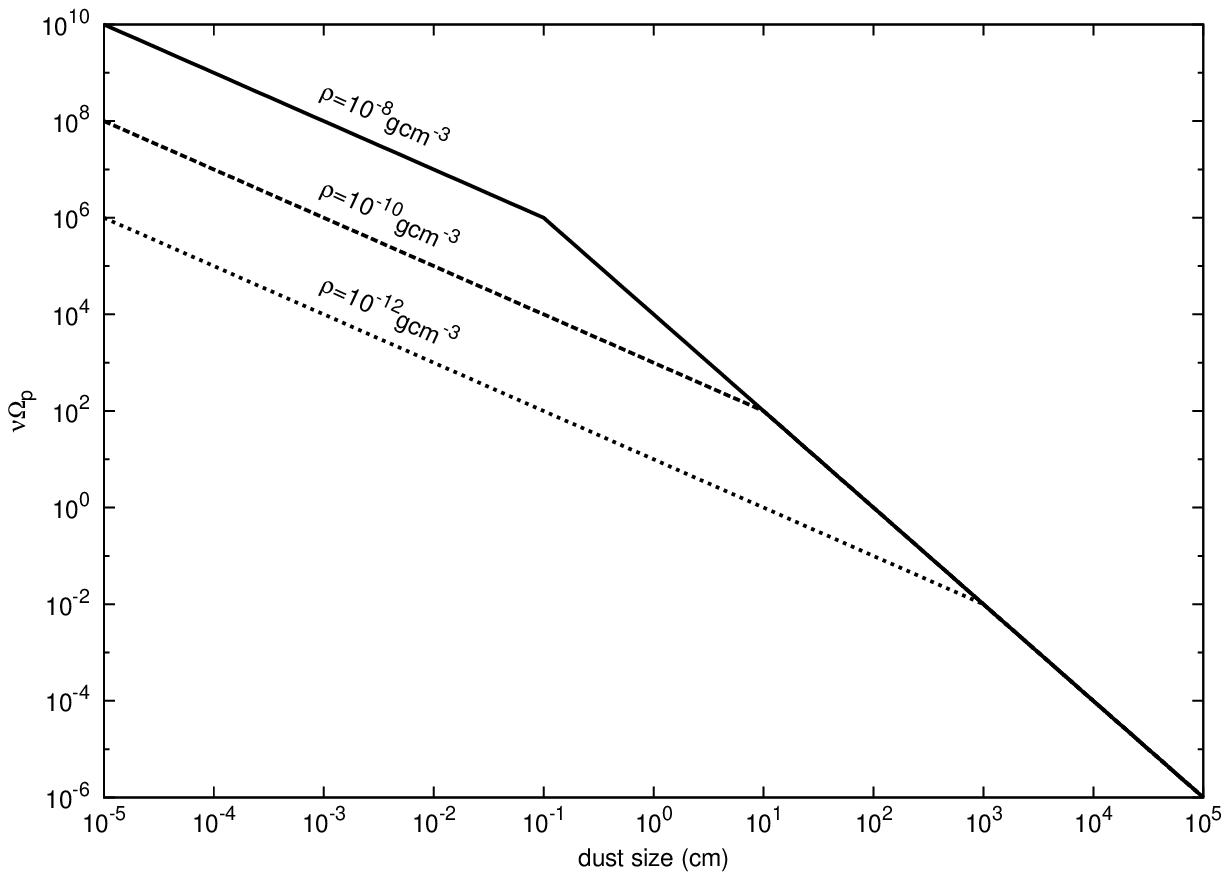}
 \caption{An example of the relationship between the dust size and the
 drag coefficient 
 $\nu$ used in our numerical calculations, given by equation
 \eqref{draglaw}.  Reciprocal of stopping time, $\nu=t_e^{-1}$ is shown
 as a function of dust size in centimeters for gas densities 
 with $\rho=10^{-8}\mathrm{g/cm}^3$, $\rho=10^{-10}\mathrm{g/cm}^3$, and
 $\rho=10^{-12}\mathrm{g/cm}^3$.}
 \label{fig:size_drag}
\end{figure}

We give a brief note on the relationship between the dust size and drag
coefficient.  In this paper, we use the constant drag coefficient $\nu$
for mathematical convenience.  The dust size $d$ and the drag
coefficient $\nu$ are related by $\nu \propto d^{-1}$ for Epstein law
and  $\nu \propto d^{-2}$ for Stokes law.  Simple relationship between
the dust size and drag coefficient may be given by
\begin{eqnarray}
 \label{draglaw}
 \dfrac{\nu}{\Omega_{\rm p}} =
  \left\{
  \begin{array}{cc}
   10^4 \left( \dfrac{\rho_0}{10^{-9} \mathrm{g/cm}^3} \right)  
    \left( \dfrac{d}{1\mathrm{cm}} \right)^{-1}   &
    \dfrac{d}{1\mathrm{cm}} < \dfrac{10^{-9} \mathrm{g/cm}^3}{\rho_0}
    \\[15pt] 
   10^4 \left( \dfrac{d}{1\mathrm{cm}} \right)^{-2}   &
    \dfrac{10^{-9} \mathrm{g/cm}^3}{\rho_0} < \dfrac{d}{1\mathrm{cm}}
  \end{array}
  \right.
\end{eqnarray}
where $\rho_0$ is the background density and $d$ is the size of the
particle.  Figure \ref{fig:size_drag} shows the relationship between
dust size and drag coefficient given by this equation for
$\rho_0=10^{-8}\mathrm{gcm}^{-3}$, $10^{-10}\mathrm{gcm}^{-3}$, and
$10^{-12}\mathrm{gcm}^{-3}$. 

The assumption of constant drag coefficient is not true for large bodies
whose drag force is proportional to 
the square of the velocity difference.  However, we expect that the
results are qualitatively similar even for large bodies.  

We neglect the effects of gravitational potential caused by gas disk.
It is a good approximation for non-self-gravitating disk as shown in
Section \ref{subsec:db_spiral}.

In the absence of the planet and the friction force, the particle
motion is the Keplerian motion around the central star.  In our setup,
this is represented by four orbital elements $(b,h,k,y_0)$,
\begin{equation}
 x(t) = b + r_{\rm p} h \cos (\Omega_{\rm p} t) + r_{\rm p} k \sin
  (\Omega_{\rm p} t)
  \label{orbit_x}
\end{equation}
\begin{equation}
 y(t) = y_0 - \dfrac{3}{2} b \Omega_{\rm p} t - 2 r_{\rm p} h \sin
  (\Omega_{\rm p} t) + 2 r_{\rm p} k \cos (\Omega_{\rm p} t) .
  \label{orbit_y}
\end{equation}
Physically, $b$ is the difference of orbital semi-major axis between the
planet and the particle, $h$ and $k$ are quantities related to the
orbital eccentricity, and $y_0$ is related to the initial azimuthal
position of the particle.

We calculate the evolution of the orbital
elements caused by the planet's gravitational force and the friction due
to the gas.  The time evolution of the osculating elements is given by
\begin{equation}
 \dot{b} = \nu r_{\rm p} \left[ h \cos (\Omega_{\rm p} t) + k \sin
			  (\Omega_{\rm p} t) \right] +
 \dfrac{2}{\Omega_{\rm p}} F_y
 \label{eq_b}
\end{equation}
\begin{equation}
 \dot{h} = -\nu h 
  - \dfrac{1}{r_{\rm p} \Omega_{\rm p}} 
  \left[ F_x \sin (\Omega_{\rm p} t) + 2 F_y \cos (\Omega_{\rm p} t)  \right]
 \label{eq_h}
\end{equation}
\begin{equation}
 \dot{k} = -\nu k 
  + \dfrac{1}{r_{\rm p} \Omega_{\rm p}} 
  \left[ F_x \cos (\Omega_{\rm p} t) - 2 F_y \sin (\Omega_{\rm p} t)
  \right] .
 \label{eq_k}
\end{equation}

\subsection{Change of Semi-Major Axis by Distant Encounter: General
  Treatment}
\label{subsec:db_formulation}

We now solve equations \eqref{eq_b}-\eqref{eq_k}.  We assume that the
particle is in circular orbit initially:
$h(t=-\infty)=k(t=-\infty)=0$.  We derive the change of orbital
semi-major axis $\Delta b=b(t=\infty)-b(t=-\infty)$ by the encounter.

The formal solutions
of equations \eqref{eq_h} and \eqref{eq_k} are given by
\begin{equation}
 h(t) = - \dfrac{1}{r_{\rm p} \Omega_{\rm p}} \int_{0}^{\infty} du 
  e^{-\nu u} \left\{ F_x(t-u) \sin [ \Omega_{\rm p}(t-u) ] + 2F_y(t-u) 
   \cos [ \Omega_{\rm p} (t-u) ]  \right\}
  \label{sol_h}
\end{equation}
\begin{equation}
 k(t) = \dfrac{1}{r_{\rm p} \Omega_{\rm p}} \int_{0}^{\infty} du 
  e^{-\nu u} \left\{ F_x(t-u) \cos [ \Omega_{\rm p}(t-u) ] - 2F_y(t-u) 
   \sin [ \Omega_{\rm p} (t-u) ]  \right\} ,
  \label{sol_k}
\end{equation}
where $F_x (t)$ denotes $F_x (x(t),y(t))$ where $(x(t),y(t))$ is
the location of the particle at time $t$ given by equations
\eqref{orbit_x} and \eqref{orbit_y} and so for $F_y$.
Substituting equations \eqref{sol_h} and \eqref{sol_k} into equation
\eqref{eq_b}, we find
\begin{eqnarray}
&  \Delta b &\equiv \int_{-\infty}^{\infty} \dot{b}(t) dt \nonumber \\ 
&& = \dfrac{\nu}{\Omega_{\rm p}}\int_{-\infty}^{\infty} dt
 \int_{0}^{\infty} du 
 e^{-\nu u} \left\{ F_x(t-u) \sin(\Omega_{\rm p} u) - 2F_y(t-u) \cos
	     (\Omega_{\rm p} u)  \right\} \nonumber \\
 && \ \ \ + \dfrac{2}{\Omega_{\rm p}}
 \int_{-\infty}^{\infty} dt F_y(t)
\end{eqnarray}
Changing the integration variable $t \to t-u$ in the first integral of
the second line and using the formula
\begin{equation}
 \int_{0}^{\infty} du e^{-\nu u} \cos (\Omega_{\rm p} u) =
  \dfrac{\nu}{\nu^2+\Omega_{\rm p}^2}
\end{equation}
and
\begin{equation}
 \int_{0}^{\infty} du e^{-\nu u} \sin (\Omega_{\rm p} u) =
  \dfrac{\Omega_{\rm p}}{\nu^2+\Omega_{\rm p}^2},
\end{equation}
we find
\begin{equation}
 \Delta b = \dfrac{\nu}{\nu^2 + \Omega_{\rm p}^2}
  \int_{-\infty}^{\infty} F_x(t)dt + 2 \dfrac{\Omega_{\rm p}}{\nu^2 +
  \Omega_{\rm p}^2} \int_{-\infty}^{\infty} F_y(t) dt
  \label{deltab_integral}
\end{equation}

This equation formally describes the amount of change of semi-major
axis.  We note that the integration with respect to $t$ must be
performed along the particle's trajectory.
The force $\vect{F}$ consists of two parts [see equation
\eqref{force_general}].
One is the gravitational force of the planet and the other is the
non-Keplerian motion of the gas velocity.  We now see these
effects separately.

\subsection{Planet Encounter}
\label{subsec:db_planet}

First, we consider the change of the semi-major axis caused by the
planet.  We assume that the trajectory of the particle is close to
circular orbit.  We approximate $F_x$ by
\begin{equation}
 F_x \sim - \mathrm{sgn}(b) \dfrac{GM_{\rm p}}{b^2}
  \dfrac{1}{( 1+(9/4)\Omega_{\rm p}t^2 )^{3/2}} .
\end{equation}
With this approximation, we obtain
\begin{equation}
 \int_{-\infty}^{\infty} F_x(t) dt = - \mathrm{sgn}(b) \dfrac{4GM_{\rm
  p}}{3b^2}.
  \label{fx_integral}
\end{equation}
For the integration of $F_y$, we can use the well-known result of
restricted three-body problem (see e.g., Goldreich \& Tremaine 1980,
H\'{e}non \& Petit 1986, or Hasegawa \& Nakazawa 1990).  The result
is 
\begin{equation}
 \int_{-\infty}^{\infty} F_y(t) dt = \dfrac{64}{243} \dfrac{G^2 M_{\rm
  p}^2}{b^5 \Omega_{\rm p}^3} \left[ K_1\left(\dfrac{2}{3}\right) + 2
			       K_0 \left(\dfrac{2}{3}\right) \right]^2,
 \label{fy_integral}
\end{equation}
where $K_0$ and $K_1$ are modified Bessel function of zeroth and first
order.  The derivation of this equation is outlined in Appendix
\ref{App:fy_int}.

Substituting equations \eqref{fx_integral} and \eqref{fy_integral} into
equation \eqref{deltab_integral}, we obtain
\begin{equation}
 \Delta b = -\mathrm{sgn}(b) 4 \dfrac{r_{\rm H}^3}{b^2} \dfrac{\nu
  \Omega_{\rm p}}{\nu^2+\Omega_{\rm p}^2} + \alpha \dfrac{r_{\rm
  H}^6}{b^5} \dfrac{\Omega_{\rm p}^2}{\nu^2 + \Omega_{\rm p}^2},
  \label{deltab_planet}
\end{equation}
where $r_{\rm H}$ is the Hill's radius of the planet defined by
\begin{equation}
 r_{\rm H} = \left( \dfrac{M_{\rm p}}{3 M_{\ast}} \right)^{1/3}
  r_{\rm p} 
\end{equation}
and $\alpha$ is a numerical factor
\begin{equation}
 \alpha \equiv \dfrac{128}{27} 
  \left[ K_1\left(\dfrac{2}{3} \right) +
	  2K_0 \left( \dfrac{2}{3} \right) \right]^2 = 30.094
\end{equation}

The second term of equation \eqref{deltab_planet} represents the effect
of gravitational scattering.  The particle and the planet tends to repel
each other by  
mutual gravitational interaction, which is a well-known result of
restricted three-body problem.  This effect remains when drag force 
vanishes ($\nu \to 0$).  When there is no drag force, the
gravitational interaction between the planet and the particle results in
the excitation of eccentricity and the difference in semi-major axes of
the two increases since Jacobi energy must be conserved in the absence
of any dissipative force.

When friction force is large, gravitational
scattering is ineffective since the gas drag enforces the particle to
move with the fluid element.  The first term of the equation
\eqref{deltab_planet} shows that the orbits of the 
planet and the particle tend to attract each other and this is most
efficient when $\Omega_{\rm p} \sim \nu$.  The intuitive
explanation of this effect is as follows.  When the particle feels the
gravitational force of the planet, it is attracted towards the planet's
position at first.  Then, when the gas drag is effective, the gas
enforces the particle to stay at the Kepler orbit at the location where
the particle is attracted, resulting in the attraction of the semi-major
axes of the particle and the planet.  If the drag force is taken into
account, the semi-major axis difference between the planet and the
particle can shrink since Jacobi energy is not necessarily conserved.

Since the attraction of the semi-major axes of the particle and
the planet in the presence of drag force represented by the first term
of equation \eqref{deltab_planet} is proportional
to $b^{-2}$ at large $b$,
 this overwhelms the effect of scattering represented by the
second term, which is proportional to $b^{-5}$.  Therefore, we conclude
that particles far away from the planet are attracted towards the planet
when the gas velocity equals Keplerian rotation velocity.

In the vicinity of the planet, it is expected that gravitational
scattering is more effective since this effect increases as $b^{-5}$.
The value of $b$ where these two effects are comparable is given by
$\Delta b=0$, that is
\begin{equation}
 |b| \sim 1.96 \left(\dfrac{\Omega_{\rm p}}{\nu} \right)^{1/3} 
  r_{\rm H}.
  \label{b_eq_grav}
\end{equation}
This indicates that particles with $\nu>\Omega_{\rm p}$ may move
towards the planet even in the absence of the global
pressure gradient.

\subsection{Non-Keplerian Rotation of Gas Disk due to Pressure Gradient}
\label{subsec:db_pressure}

In this subsection, we consider the change of the semi-major axis caused
by the effect of non-Keplerian rotation of the gas, which is due to the
presence of global pressure gradient.  The result of this
section is already derived by previous studies such as Adachi et
al. (1976) or Weidenschilling (1977), but we briefly show the results in
this formulation for completeness.

We parameterize the degree of non-Keplerian rotation by $\eta$ and write
the velocity difference $\delta \vect{v}_{\rm g}$ by
\begin{equation}
 \delta \vect{v}_{\rm g} = \eta v_{\rm p} \vect{e}_{y} = \mathrm{const},
  \label{parameter_nonkepler}
\end{equation}
where $v_{\rm p}$ is the rotation velocity of the planet around the
central star, $v_{\rm p} = r_{\rm p} \Omega_{\rm p}$.  The parameter
$\eta$ is non-dimensional and its magnitude is the order of the square
of the disk aspect ratio.  This is negative for the disk with negative
pressure gradient.

Since $\delta \vect{v}_{\rm g}$ is constant, it is easy to integrate
\eqref{deltab_integral} to obtain
\begin{equation}
 \Delta b = 2 \eta v_{\rm p} T \dfrac{\nu \Omega_{\rm p}}{\nu^2 +
  \Omega_{\rm p}^2},
  \label{deltab_nonkepler}
\end{equation}
where $T$ is the time taken for the particle to cross the box.  As
is well-known, this effect is most efficient for the particle with
$\nu\sim\Omega_{\rm p}$.

\subsection{Steady Accretion Flow}

In this subsection, we consider the change of the semi-major axis caused
by the steady accretion (or deccretion) flow.  This may be modeled by
the constant axisymmetric radial velocity $\delta v_x$.  We parameterize
the radial velocity by non-dimensional factor $\zeta$ as follows,
\begin{equation}
 \delta \vect{v}_{\rm g} = \zeta v_{\rm p} \vect{e}_x = \mathrm{const}.
  \label{parameter_accretion}
\end{equation}
The parameter $\zeta$ denotes the ratio between the radial flow velocity
and the Kepler velocity.  If the dissipation of the gaseous nebula is
due to the gas accretion, the sign of $\zeta$ is negative and its
magnitude is the order of $10^{-6}$, which
is indicated from observation (see, e.g., Haisch et al. 2001).  
Drag force due to the gas is given by $\nu \delta \vect{v}_{\rm g}$ and
therefore, using equation \eqref{deltab_integral}, we
find
\begin{equation}
 \Delta b = \zeta v_{\rm p} T \dfrac{\nu^2}{\nu^2+\Omega_{\rm p}^2} .
  \label{deltab_accretion}
\end{equation}
We note that the steady accretion flow effect is important for small
particles, which have large drag coefficient $\nu$, since such particles
move in the same way as the gas flow.

\subsection{Non-Keplerian Rotation of Gas due to the Presence of the
  Planet}
\label{subsec:db_spiral}

In this section, we consider the change of the orbital semi-major axis
of the particle caused by the deviation of the gas velocity from
Keplerian rotation velocity due to the presence of the planet. 
In this section, $\delta \vect{v}_{\rm g}$ is caused by the
gravitational interaction between gas and the embedded planet, in
contrast to the previous sections.  We show that only axisymmetric
structure of the disk causes the change of the semi-major axis of the
particle.  

In order to calculate the velocity perturbation caused by the planet, we
solve vertically averaged Euler equations using local shearing-sheet
approximation:
\begin{eqnarray}
\label{EoC_hydro}
& \dfrac{\partial \Sigma}{\partial t} + \nabla \cdot (\Sigma \,
\vect{v}) =  0,\\
\label{EoM_hydro}
& \dfrac{\partial \vect{v}}{\partial t} + (\vect{v} \cdot \nabla)
 \vect{v} = 3\Omega_{\rm p}^2 x \vect{e}_x 
- \dfrac{c^2}{\Sigma} \nabla \Sigma 
 - \nabla \psi_{\rm p} - 2 \Omega_{\rm p} (\vect{e}_z \times \vect{v}), 
\end{eqnarray}
where $\Sigma$ denotes the surface density, $\vect{v}$ is the gas
velocity, $c$ is the sound speed, and $\Omega_{\rm p}$ is the Keplerian
angular velocity of the planet.  We assume an isothermal disk, where $c$
is constant, for simplicity.  
The gravitational potential of the
planet, $\psi_{\rm p}$ is given by
\begin{equation}
 \psi_{\rm p} = -\dfrac{GM_{\rm p}}{\sqrt{x^2+y^2+\epsilon^2}},
\end{equation}
where $\epsilon$ is the softening parameter.
Taking the rotation of the equation of motion, we obtain the equation
for vorticity,
\begin{equation}
 \left( \dfrac{\partial}{\partial t} + \vect{v}\cdot\nabla  \right)
 \left[ \dfrac{1}{\Sigma} \left\{ \dfrac{\partial v_y}{\partial x}
		   - \dfrac{\partial v_x}{\partial y} 
		   + 2\Omega_{\rm p} \right\} \right] = 0 .
 \label{EoV_hydro}
\end{equation}

For small mass planets we are interested in this paper, it is possible
to calculate the velocity perturbation by linear analysis.  However, as
shown later, it is necessary to calculate the flow up to the second
order in order to obtain the correct results for the particle motion.
We assume that background surface density $\Sigma_0$ is constant and
background gas is rotating with Kepler velocity 
$\vect{v}_0=-(3/2)\Omega_{\rm p}x\vect{e}_y$.  We neglect the effect of
the global pressure gradient or steady mass accretion since it is
calculated separately in the previous subsections.

The first order perturbation of the flow is caused by the planet
gravity.  We denote the first order perturbation of physical variables
by superscript $(1)$.  The linearization of equations
\eqref{EoC_hydro} and \eqref{EoM_hydro} are
\begin{eqnarray}
\label{EoC_hydro_lin}
& \left(\dfrac{\partial}{\partial t} - \dfrac{3}{2}
  \Omega_{\rm p} x \dfrac{\partial}{\partial y} \right) \dfrac{\delta
 \Sigma^{(1)}}{\Sigma_0} + \dfrac{\partial}{\partial x} \delta v_x^{(1)}
 + \dfrac{\partial}{\partial y} \delta v_y^{(1)} = 0 \\
\label{EoMx_hydro_lin}
& \left(\dfrac{\partial}{\partial t} - \dfrac{3}{2}
  \Omega_{\rm p} x \dfrac{\partial}{\partial y} \right) \delta v_x^{(1)}
- 2 \Omega_{\rm p} \delta v_y^{(1)} = -c^2 \dfrac{\partial}{\partial x}
\dfrac{\delta \Sigma^{(1)}}{\Sigma_0} 
- \dfrac{\partial}{\partial x} \psi_{\rm p} \\
\label{EoMy_hydro_lin}
& \left(\dfrac{\partial}{\partial t} - \dfrac{3}{2}
  \Omega_{\rm p} x \dfrac{\partial}{\partial y} \right) \delta v_y^{(1)}
+ \dfrac{1}{2} \Omega_{\rm p} \delta v_x^{(1)} 
 = -c^2 \dfrac{\partial}{\partial y} \dfrac{\delta
 \Sigma^{(1)}}{\Sigma_0} 
 - \dfrac{\partial}{\partial y} \psi_{\rm p} .
\end{eqnarray}
Assuming that vorticity is not created by the formation of the planet,
we have
\begin{equation}
 \dfrac{\partial}{\partial y} \delta v_x^{(1)} -
  \dfrac{\partial}{\partial x} 
  \delta v_y^{(1)} + \dfrac{1}{2} \Omega_{\rm p} \dfrac{\delta
  \Sigma^{(1)}}{\Sigma_0} =0 ,
  \label{vorticity}
\end{equation}
see, e.g., Narayan et al. (1987).
Note that $\Sigma_0$ appears only in the form of 
$\delta \Sigma^{(1)} / \Sigma_0$
and therefore, this value itself does not depend on the background
surface density.
The source term of these equations is given by planet potential,
$\psi_{\rm p}$ so the perturbation is proportional to the planet mass
$M_{\rm p}$.  For homogeneous ($\psi_{\rm p}=0$) equations, only
dimensional parameters are $c$ and $\Omega_{\rm p}$.  Taking 
$\Omega_{\rm p}^{-1}$ as a unit of time and $H=c/\Omega_{\rm p}$ as a
unit of length, homogeneous equations become scale free. 
Since the amplitude of the perturbation is proportional to the source
term, which is $\psi_{\rm p}$ in this set of equations, the
perturbation amplitude scales with $\psi_{\rm p}/c^2$, in other words, 
\begin{equation}
 \dfrac{\delta \Sigma^{(1)}}{\Sigma_0} \propto \dfrac{GM_{\rm p}}{H c^2} 
  \label{amp_order}
\end{equation}
as shown by Tanaka et al. (2002).  Once we know the perturbation amplitude
of a specific value of $GM_{\rm p}/Hc^2$, we can easily obtain the actual
amplitude of density perturbation $\delta \Sigma$ for different
background density, protoplanet mass, sound speed and so on.  Linear
calculation shows that the coefficient of proportionality of equation
\eqref{amp_order} is the order of unity.

The first order perturbation becomes a source of second order
perturbations through non-linear terms of basic equations.  We denote
the second order perturbation by superscript $(2)$.  The set of second
order perturbation equations is
\begin{eqnarray}
\label{EoC_hydro_2nd}
& \left(\dfrac{\partial}{\partial t} - \dfrac{3}{2}
  \Omega_{\rm p} x \dfrac{\partial}{\partial y} \right) \dfrac{\delta
 \Sigma^{(2)}}{\Sigma_0} + \dfrac{\partial}{\partial x} \delta v_x^{(2)}
 + \dfrac{\partial}{\partial y} \delta v_y^{(2)} \nonumber \\
& \ \ \ \ \ \ \ \ \ \ 
 = - \left[ \dfrac{\partial}{\partial x} \left( \delta \Sigma^{(1)}
					  \delta v_x^{(1)} \right)
  + \dfrac{\partial}{\partial y} \left( \delta \Sigma^{(1)} \delta
			       v_y^{(1)} \right) 
  \right] \\
\label{EoMx_hydro_2nd}
& \left(\dfrac{\partial}{\partial t} - \dfrac{3}{2}
  \Omega_{\rm p} x \dfrac{\partial}{\partial y} \right) \delta v_x^{(2)}
- 2 \Omega_{\rm p} \delta v_y^{(2)} + c^2 \dfrac{\partial}{\partial x}
\dfrac{\delta \Sigma^{(2)}}{\Sigma_0}  \nonumber \\
& \ \ \ \ \ \ \ \ \ \ 
 = c^2 \dfrac{\delta \Sigma^{(1)}}{\Sigma_0} \dfrac{\partial}{\partial x}
\dfrac{\delta \Sigma^{(1)}}{\Sigma_0} - \delta v_x^{(1)}
\dfrac{\partial}{\partial x} \delta v_x^{(1)} - \delta v_y^{(1)}
\dfrac{\partial}{\partial y} \delta v_x^{(1)} \\
\label{EoMy_hydro_2nd}
& \left(\dfrac{\partial}{\partial t} - \dfrac{3}{2}
  \Omega_{\rm p} x \dfrac{\partial}{\partial y} \right) \delta v_y^{(2)}
+ \dfrac{1}{2} \Omega_{\rm p} \delta v_x^{(2)} 
 + c^2 \dfrac{\partial}{\partial y} \dfrac{\delta
 \Sigma^{(2)}}{\Sigma_0}  \nonumber \\
& \ \ \ \ \ \ \ \ \ \ 
 = c^2 \dfrac{\delta \Sigma^{(1)}}{\Sigma_0} \dfrac{\partial}{\partial y}
\dfrac{\delta \Sigma^{(1)}}{\Sigma_0} - \delta v_x^{(1)}
\dfrac{\partial}{\partial x} \delta v_y^{(1)} - \delta v_y^{(1)}
\dfrac{\partial}{\partial y} \delta v_y^{(1)} .
\end{eqnarray}
The second-order vorticity equation is
\begin{equation}
 \dfrac{\partial}{\partial y} \delta v_x^{(2)} -
  \dfrac{\partial}{\partial x} 
  \delta v_y^{(2)} + \dfrac{1}{2} \Omega_{\rm p} \dfrac{\delta
  \Sigma^{(2)}}{\Sigma_0} =
 - \dfrac{\delta \Sigma^{(1)}}{\Sigma_0} 
 \left[ \dfrac{\partial}{\partial x} \delta v_y^{(1)} -
  \dfrac{\partial}{\partial y} \delta v_x^{(1)}
  \right] + \dfrac{1}{4} \left( \dfrac{\delta \Sigma^{(1)}}{\Sigma_0}
			 \right)^2 .
\label{vorticity_2nd}
\end{equation}
Since the first order perturbation scales with $GM_{\rm p}/Hc^2$ (see
equation \eqref{amp_order}), it is clear that the second order
perturbation scales with $(GM_{\rm p}/Hc^2)^2$. 
We assume that the perturbation is stationary with respect to the planet
so in the corotating frame with the planet considered in this
subsection, $\partial/\partial t = 0$. 

%%%%%%%%%%%%%moved to later section in revision%%%%%%%%%
% In order to find the spiral
%pattern produced by the planet, we numerically integrate 
%equations \eqref{EoC_hydro_lin}-\eqref{EoMy_hydro_lin} using the Fourier
%transform methods presented by Goodman and Rafikov (2001).  This method
%is particularly useful in calculating the gravitational potential
%produced by the density fluctuation. 
%We fix the spatial resolution
% (therefore, the maximum wave number in Fourier space)
%$(\Delta x/H, \Delta y/H)$ to be 
%$(9.8 \times 10^{-3}, 7.8 \times 10^{-2} )$, which is enough to resolve
% the effective Lindblad resonance (see e.g., Artymowicz 1993).  We fix
% the box size in $y$-direction such that $-80H<y<80H$ and impose a
% periodic boundary condition.  For $x$-direction, the
% resolution in Fourier space in $k_x$ direction is varied from $10^4$ to
% $10^6$ according to $k_y$ in order 
% to resolve very fast oscillation at $k_x \gg k_y$.  We also use the
% linear window function given by Goodman and Rafikov (2001).
% We set softening parameter $\epsilon$ to be $0.01H$.   This numerical
% data is used in integrating the particle orbits in section
% \ref{numerical}. 
%%%%%%%%%%%%%moved to later section in revision end%%%%%%%%%

We now see equations \eqref{EoC_hydro_lin}-\eqref{EoMy_hydro_lin}
in more detail to find an analytic expression of $\Delta b$.  We first
give an order-of-magnitude discussion to show that the gravitational
force produced by the density fluctuation is small compared to the
gravitational force by the planet.  We expect that the order of
magnitude of the gravitational force by the spiral density fluctuation
is 
\begin{equation}
 F_{\rm spiral} \sim \dfrac{G \delta \Sigma H^2}{H^2},
\end{equation} 
since the characteristic length scale of the spiral is of the order
of the scale height.  On the other hand, the gravitational force by the
planet exerted on the particle at the distance of the order of the scale
height is 
\begin{equation}
 F_{\rm planet} \sim \dfrac{GM_{\rm p}}{H^2}.
\end{equation}
Using equation \eqref{amp_order}, we expect that these two forces are
related by
\begin{equation}
 \dfrac{F_{\rm spiral}}{F_{\rm planet}} \sim 
  \dfrac{G\Sigma_0}{c\Omega_{\rm p}} \equiv Q^{-1},
\end{equation}
where $Q$ is Toomre's $Q$ parameter of the disk.  Since protoplanetary
disks in planet forming phase are
expected to be gravitationally stable in general, $Q \gg 1$.
Therefore, we conclude that the gravitational interaction between the
particle and the disk is negligible.  We later confirm this
numerically.  In this section, we consider only
the effect of the velocity fluctuation of the gas disk in the presence
of the planet.

\subsubsection{First-Order Axisymmetric Mode}

\begin{figure}
 \plotone{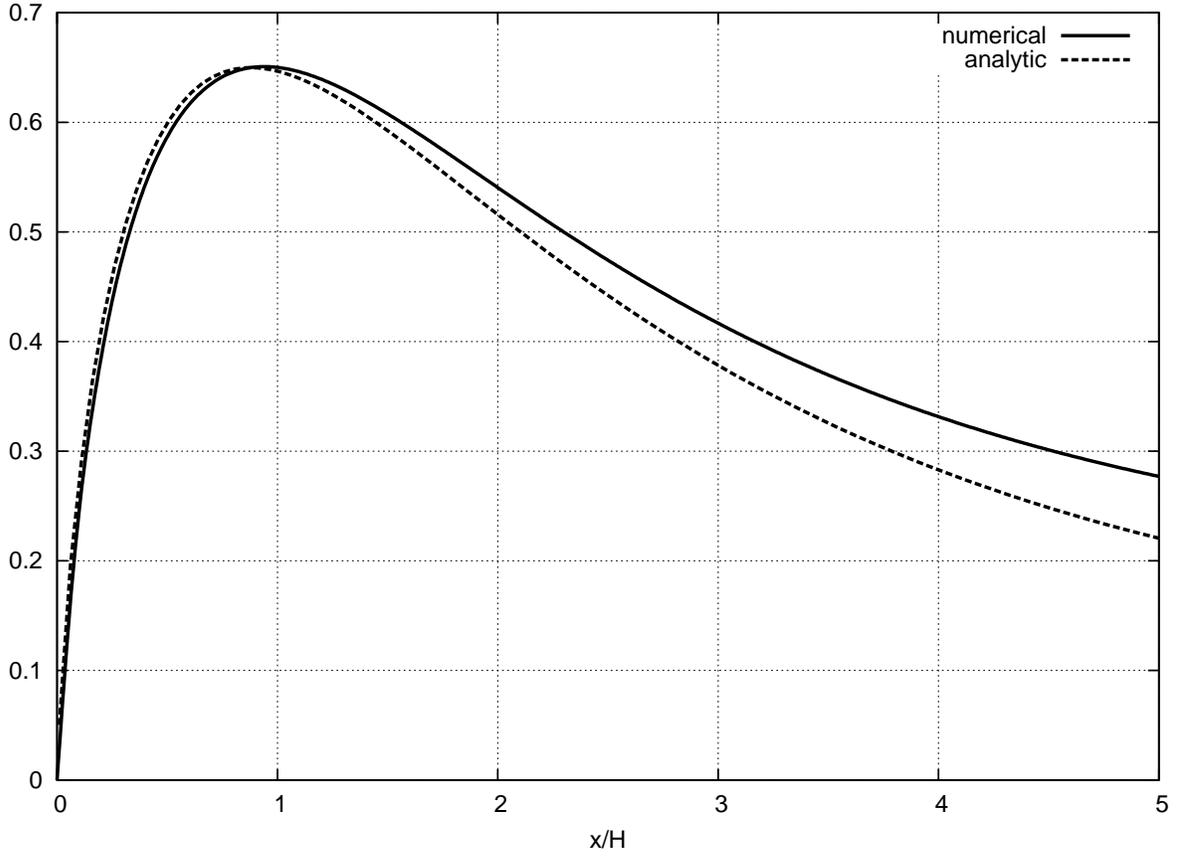}
 \caption{Comparison between the azimuthally averaged profile of 
$L_y \delta v_y/c$ obtained by the
 numerical calculation (solid line) and analytic expression
 \eqref{dvy_asym} (dotted line).  
The results of the numerical calculation is normalized by 
$GM_{\rm p}/Hc^2$.}
 \label{fig:dvy_asym_profile}
\end{figure}

We first consider axisymmetric modes, or azimuthally averaged
quantity.  For any perturbation quantities $\delta f$, we denote azimuthal
average by bars,
\begin{equation}
 \overline{\delta f}(x) = \dfrac{1}{L_y} \int_{-L_y/2}^{L_y/2} \delta
  f (x,y) dy .
\end{equation}
Using Euler equations \eqref{EoC_hydro_lin}-\eqref{EoMy_hydro_lin} and
vorticity equation \eqref{vorticity}, we can derive
\begin{eqnarray}
\label{dvx_asym}
& \overline{\delta v_x^{(1)}} = 0, \\
\label{dsigma_dvy}
& \dfrac{\overline{\delta \Sigma^{(1)}}}{\Sigma_0} =
\dfrac{2}{\Omega_{\rm p}} 
\dfrac{d}{dx} \overline{\delta v_y^{(1)}}, \\
\label{dvy_asym_eqn} 
& \dfrac{d^2}{dx^2} \overline{\delta v_y^{(1)}} - \dfrac{\Omega_{\rm 
 p}^2}{c^2} \overline{\delta v_y^{(1)}} = -\dfrac{\Omega_{\rm p}}{2c^2} 
 \dfrac{d\overline{\psi}}{dx}.
\end{eqnarray}
For $x\ll L_y$, azimuthal average of $\partial \psi/\partial x$ is given
by 
\begin{equation}
 \dfrac{d\overline{\psi}}{dx} \sim 
  \dfrac{1}{L_y}\int_{-\infty}^{\infty} \dfrac{GM_{\rm
  p}x}{\left(x^2+y^2\right)^{3/2}} dy = \dfrac{2GM_{\rm p}}{L_y x},
\end{equation}
where the integration range is extended to infinity.
The appropriate boudary condition is such that the perturbation vanishes
for $|x|\to\infty$.  Since homogeneous solution of equation
\eqref{dvy_asym_eqn} is $\exp[\pm x/H]$, Green's function of the
differential operater of equation \eqref{dvy_asym_eqn} is
\begin{equation}
 G(x;x^{\prime}) = -\dfrac{H}{2} 
  \left[ e^{-(x-x^{\prime})/H} \theta(x-x^{\prime})
   +  e^{(x-x^{\prime})/H} \theta(x^{\prime}-x)  \right] ,
\end{equation}
where $\theta(x)$ is step function.
Therefore, the solution
of equation \eqref{dvy_asym_eqn} is given by
\begin{equation}
 \overline{\delta v_y^{(1)}} = \dfrac{H^2 \Omega_{\rm p}}{2L_y}
  \dfrac{GM_{\rm p}}{Hc^2} 
  \left[ e^{-(x/H)} \mathrm{Ei}\left( \dfrac{x}{H} \right) 
   - e^{x/H} \mathrm{Ei}\left( -\dfrac{x}{H} \right) \right] .
  \label{dvy_asym}
\end{equation}
Using equation \eqref{dsigma_dvy}, the surface density perturbation is
given by 
\begin{equation}
 \dfrac{\overline{\delta \Sigma^{(1)}}}{\Sigma_0} = - \dfrac{H}{L_y} 
  \dfrac{GM_{\rm p}}{Hc^2}
  \left[ e^{-(x/H)} \mathrm{Ei}\left( \dfrac{x}{H} \right) 
   + e^{x/H} \mathrm{Ei}\left( -\dfrac{x}{H} \right) \right],
  \label{dens_asym}
\end{equation}
where $\mathrm{Ei}(x)$ denotes exponential integral.  Equations
\eqref{dvy_asym} and \eqref{dens_asym} are valid for locations 
$x\ll L_y$.  Figure \ref{fig:dvy_asym_profile} shows the profile of
$L_y\overline{\delta v_y^{(1)}}(x)/c$ normalized by $(GM_{\rm p}/Hc^2)$
and compares this with analytic expression \eqref{dvy_asym}.  Difference 
between numerical and analytic results is about $25 \%$ at $x\sim 5H$.
We have checked that this error significantly decreases when box size in
$x$-direction is smaller than that of $y$-axis.

\subsubsection{Second-Order Axisymmetric Mode and Path Line of the Flow}

The second-order axisymmetric modes are obtained by equations
\eqref{EoMx_hydro_2nd}, \eqref{EoMy_hydro_2nd}, and
\eqref{vorticity_2nd}.  We note that equation of continuity
\eqref{EoC_hydro_2nd} and the $y$-component of equation of motion
\eqref{EoMy_hydro_2nd} is not independent for axisymmetric modes.
We use only $\overline{\delta v_x^{(2)}}$ later.  This is given by
\begin{equation}
 \overline{\delta v_x^{(2)}} = - \dfrac{2}{\Omega_{\rm p}}
 \overline{ \delta v_x^{(1)} \dfrac{\partial}{\partial x} \delta
 v_y^{(1)}}.
 \label{dvx_2nd}
\end{equation}

We then consider the path line of the flow.  This gives the motion of
the fluid element and therefore gives the motion of the particle with
$\nu\to\infty$.  Since we are interested in the change of the orbital
semi-major axis of the particle, we consider the motion of the fluid
element in the $x$-direction.  Path line is given by
\begin{eqnarray}
\label{pathline_x}
& \dfrac{dx}{dt} = \delta v_x (x(t),y(t))  \\
\label{pathline_y}
& \dfrac{dy}{dt} = v_c + \delta v_y(x(t),y(t))
\end{eqnarray}
where $v_c$ is the unperturbed velocity
\begin{equation}
 v_c = -\dfrac{3}{2} \Omega_p x
\end{equation}
and $\delta v_x$ and $\delta v_y$ include both first and second order
perturbations; for instance, 
$\delta v_x = \delta v_x^{(1)} + \delta v_x^{(2)}$.

The path line of the unperturbed flow is given by 
\begin{eqnarray}
& x_c(t) = b = \mathrm{const} \\
& y_c(t) = \dfrac{L_y}{2}  -\dfrac{3}{2} b \Omega_{\rm p} t
\end{eqnarray}
where we assume that the fluid element is at $(b,L_y/2)$ at $t=0$.  This
is simply a Keplerian circular motion.  We
denote the time when the particle reaches the other end of the box in
the unperturbed flow by $t=T$, where
\begin{equation}
 T = \dfrac{L_y}{(3/2) b \Omega_{\rm p}} .
\end{equation}

We solve equations \eqref{pathline_x} and \eqref{pathline_y} using
perturbation methods and obtain the change of $x$-coordinate,
 $\Delta x$, of the fluid element after it crosses the box
\begin{equation}
 \Delta x \equiv x(T)-x(0) = \int_0^T dt \delta v_x(x(t), y(t)) .
  \label{pathline_Dx_def}
\end{equation}
We divide the motion of the fluid element into the unperturbed motion
and perturbation
\begin{eqnarray}
& x(t) = x_c + \delta x(t) \\ 
& y(t) = y_c(t) + \delta y(t).
\end{eqnarray}
Expanding right hand side of equation
\eqref{pathline_Dx_def} upto the first order of $\delta x$ and 
$\delta y$ and using equations \eqref{pathline_x} and
\eqref{pathline_y}, we obtain
\begin{eqnarray}
& \Delta x =& \int_0^T \delta v_x^{(2)} (x_c, y_c(t)) dt
  + \int_0^T dt_1 \dfrac{\partial \delta v_x^{(1)}}{\partial x}
  (x_c,y_c(t_1))  
    \int_0^{t_1} dt_2 \delta v_x^{(1)} (x_c,y_c(t))  \nonumber \\ 
&&  + \int_0^T dt_1 \dfrac{\partial \delta v_x^{(1)}}{\partial y}
  (x_c,y_c(t_1))  
    \int_0^{t_1} dt_2 \delta v_y^{(1)} (x_c,y_c(t))
\label{Dx_calc}
\end{eqnarray}
where we have used $\overline{\delta v_x^{(1)}}=0$ and take the terms up
to the second order.  Using the first order vorticity equation
\eqref{vorticity} and equation of continuity, it is possible to show
that the right hand side of equation \eqref{Dx_calc} vanishes, see
Appendix \ref{app:pathline}.  Therefore, we conclude that fluid element
returns to the original radial position after crossing the spiral
density wave,
\begin{equation}
 \Delta x = 0 .
  \label{Dx_vanish}
\end{equation}
We note that second-order perturbation is essential to derive this
conclusion and therefore, it is necessary to consider the fluid motion
upto the second order in finding the correct motion of the particles in
a disk.

The fact that the fluid element does not move in the radial direction
indicates that there is no radial mass flux up to second order
perturbation.  It is possible to show 
this directly by calculating $\Sigma v_x$,
\begin{eqnarray}
& \displaystyle{\int} dy \Sigma v_x 
 &= \int dy \Sigma_0 \delta v_x^{(2)} + \Sigma_0
 \int dy \dfrac{\delta \Sigma^{(1)}}{\Sigma_0} \delta v_x^{(1)}
 \nonumber \\
&  &= -\dfrac{2}{\Omega_{\rm p}} \Sigma_0 \int dy \delta v_x^{(1)}
 \dfrac{\partial}{\partial x} \delta v_y^{(1)}
 + \Sigma_0 \int dy \dfrac{\delta \Sigma^{(1)}}{\Sigma_0} \delta
 v_x^{(1)} \nonumber \\
&  &= -\dfrac{1}{\Omega_{\rm p}}\Sigma_0 \int dy
 \dfrac{\partial}{\partial y} \left( \delta v_x^{(1)} \right)^2
			       \nonumber \\
& &= 0,
\label{massflux_vanish}
\end{eqnarray}
where we have used equation \eqref{dvx_2nd} in the second line,
vorticity equation \eqref{vorticity} in the third line, and periodic
boundary condition in the last equality
%%%%%%%%footnote added in revision%%%%%%%%
\footnote{It is actually possible to show that, in the shearing-sheet 
 approximation, the radial mass flux
vanishes for all orders of perturbation in the same manner as presented
here.}.
%%%%%%%%footnote added in revision end%%%%%%%%
This is in contrast to the case of the sound wave propagating in a
static, homogeneous medium, which carries the linear momentum and
therefore mass flux is present (Landau and Lifshitz 1959).  We
demonstrate this in Appendix \ref{app:soundwave}.

\subsubsection{$\Delta b$ Caused by Spiral Density Wave}

We are now in the position to calculate $\Delta b$ due to the structure
of the gas disk modified by the planet's gravity.  Using equation
\eqref{deltab_integral} and drag force is 
given by $\nu \delta \vect{v}_{\rm g}$, we have
\begin{equation}
 \Delta b = \dfrac{\nu^2}{\nu^2 + \Omega_{\rm p}^2}
  \int_{-\infty}^{\infty} \delta v_x \left(x(t),y(t) \right) dt +
  \dfrac{2\nu\Omega_{\rm p}}{\nu^2+\Omega_{\rm p}^2}
  \int_{-\infty}^{\infty} \delta v_y \left( x(t),y(t) \right) dt.
  \label{deltab_deltav_int}
\end{equation}
The integration must be performed along the path of the particle.  The
first term of equation \eqref{deltab_deltav_int} dominates when
$\nu\to\infty$.  In the perfect coupling limit, since the particle
traces the flow of the fluid element, this integral should vanish.  We
expect that this integral always remains small and set the first term of
equation \eqref{deltab_deltav_int} to 
be zero.  We expect that this approximation is valid since the second
term that is proportional to  
$\nu\Omega_{\rm p}/(\nu^2+\Omega_{\rm p}^2)$ dominates $\Delta b$ when
$\nu$ is small.  We later confirm this numerically.

We then consider the second term of equation
\eqref{deltab_deltav_int}.  The leading order of this integration is
obtained by approximating that the particle orbit is circular.  Since
$\overline{\delta v_y^{(1)}}\neq 0$, it is enough to consider this
contribution.  Therefore, we arrive at the following expression of the
change of orbital semi-major axis caused by the spiral structure of the
gas
\begin{equation}
 \Delta b = \mathrm{sgn}(b) \dfrac{4}{3} \dfrac{1}{b\Omega_{\rm p}} 
 \dfrac{\nu \Omega_{\rm p}}{\nu^2+\Omega_{\rm p}^2} 
 L_y \overline{\delta v_y^{(1)}},
  \label{deltab_spiral}
\end{equation} 
where $\overline{\delta v_y^{(1)}}$ is given by equation \eqref{dvy_asym}.

We note that in deriving equation \eqref{deltab_spiral}, we have
repeatedly used the assumption of no vorticity source, equation
\eqref{vorticity}.  In the presence of vorticity source,
 $\overline{\delta v_y^{(1)}}$ may be different and the mass flux may be
 present.

\subsection{Analytic Expression of the Change of the Orbital Semi-Major
  Axis}
  \label{subsec:db_anal}

Adding equations \eqref{deltab_planet}, \eqref{deltab_nonkepler},
\eqref{deltab_accretion}, and \eqref{deltab_spiral}, we
obtain the expression for the secular evolution of the orbital
semi-major axis of the particle.  We write in the
form of the rate of the change by dividing $\Delta b$ by
$T=L_y/(3/2)\Omega_{\rm p} |b|$, where $L_y$ is the box size of the
$y$-direction of the coordinate system considered.  
Time $T$ may be interpreted as the time interval between successive
conjunctions between the particle and the planet.
The result is
\begin{eqnarray}
& \dfrac{\Delta b}{T} =& 2\eta v_{\rm p} \dfrac{\nu \Omega_{\rm p}}{\nu^2
  + \Omega_{\rm p}^2} 
  + \zeta v_{\rm p} \dfrac{\nu^2}{\nu^2+\Omega_{\rm p}^2} \nonumber \\
&&  - \mathrm{sgn}(b) \dfrac{4}{T} \dfrac{r_{\rm H}^3}{b^2} \dfrac{\nu
  \Omega_{\rm p}}{\nu^2 + \Omega_{\rm p}^2}
  + \dfrac{\alpha}{T} \dfrac{r_{\rm H}^6}{b^5} 
  \dfrac{\Omega_{\rm p}^2}{\nu^2 + \Omega_{\rm p}^2} \nonumber \\ 
&& + \mathrm{sgn}(b) \dfrac{2}{T} \dfrac{r_{\rm H}^3}{bH} 
  \left[ e^{-(b/H)} \mathrm{Ei}\left( \dfrac{b}{H} \right) 
   - e^{b/H} \mathrm{Ei}\left( -\dfrac{b}{H} \right) \right]
  \dfrac{\nu \Omega_{\rm p}}{\nu^2+\Omega_{\rm p}^2},
  \label{deltab_final}
\end{eqnarray}
where $r_{\rm H}$ is Hill's radius of the planet, the definitions of
$\eta$ and $\zeta$ are given by equations \eqref{parameter_nonkepler}
and \eqref{parameter_accretion} respectively, and the value of $\alpha$
is $30.094$.

If the planet mass is sufficiently small, the leading contribution comes
from the first term, which is due to the global pressure gradient
exerted on the gas disk.  Comparing the order of magnitude of the first
and the third term of equation \eqref{deltab_final}, the third term
becomes dominant for particles with
\begin{equation}
 \dfrac{|b|}{H} \lesssim \left( \dfrac{r_{\rm H}}{H} \right)^{3/2},
\end{equation}
where we have assumed $L_y\sim r_{\rm p}$, $\eta\sim (H/r_{\rm p})^2$
and all the numerical coefficients are neglected.  For low mass planets
which does not form a gap in the gas disk, Hill's radius is generally
smaller than the scale height.  Therefore, the motion of the particle is 
predominated by the effect of global pressure gradient.  However, for
large planets with mass of order Jupiter, the effect of the global
pressure gradient is small compared to the gravitational interaction
between the planet, and equation \eqref{deltab_final} indicates that if
the profile of the gas disk is neglected, the particles may be attracted
towards the planet.  This seems to be consistent with what is found by
Paardekooper (2007).  He found that when axisymmetric gas disk profile
is considered, plants with the mass of the order of Jupiter would
accrete the 
particles with stopping time comparable to Kepler time, while accretion
of such particles around planets with 0.1 times Jupiter mass would be
inefficient (see figures 7 and 11 of Paardekooper 2007).  When global
pressure gradient of gas disk drives the 
particles to migrate towards the central star, a part of particles with  
$\nu\sim\Omega_{\rm p}$ may pass the orbit of the planet without
accumulating onto it, since the particle migration rate
is very fast and the perturbation of the orbital elements of the
particles is not related to the planet location (Kary et al. 1993).  If
the third term of equation
\eqref{deltab_final} dominates over the global pressure effect, on the
other hand, particles will accumulate onto the planet since their
orbital elements are strongly perturbed at the location of the planet. 

\begin{figure}
 \plotone{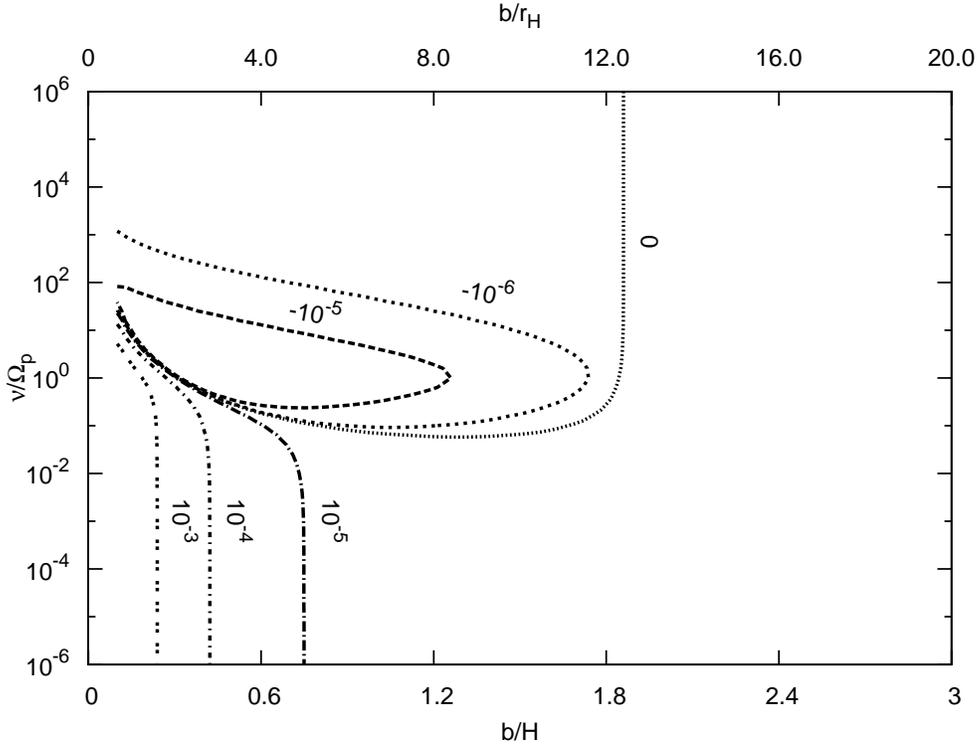}
 \caption{Contour plot of $\Delta b/T$ given by equation
 \eqref{deltab_final} normalized by $H\Omega_{\rm p}$ for a planet with 
$GM_{\rm p}/Hc^2=10^{-2}$.  The horizontal axis shows the distance from
 the planet and the vertical axis shows the drag coefficient of the
 particle. Global pressure gradient and steady mass accretion is
 neglected ($\eta=\zeta=0$).  We note that our analytic approach is
 limited by assumptions of local approximation, initially circular orbit
 of the particle, and distant encounter.  This figure has only
 qualitative meanings for particles with 
 $\nu/\Omega_{\rm p}\lesssim10^{-2}$, with initial semi-major axis
 difference $b \lesssim 3r_{\rm H}$ for 
 $\nu/\Omega_{\rm p} \lesssim 1-10$, or with 
 $b \lesssim r_{\rm H}$ for $\nu/\Omega_{\rm p} \gtrsim 10$.  Detailed
 discussions in the limitation of analytic calculations are given in
 Sections \ref{subsec:limitation_local} and \ref{global}, and are
 summarized in Section \ref{summary}.}
 \label{fig:db_contour}
\end{figure}

If the disk model with flat pressure profile is 
considered, the behavior of the particle around the planet is
complicated. Figure \ref{fig:db_contour} shows the contour plot of equation
\eqref{deltab_final} for the planet mass $GM_{\rm p}/Hc^2=10^{-2}$,
flat pressure profile ($\eta=0$), and no mass accretion onto the central
star ($\zeta=0$).  
For small particles ($\nu\to\infty$), there is no
secular motion of 
particles since all the terms of equation \eqref{deltab_final} vanishes
if $\zeta=0$.  The motion of small particles is determined by the
strength of accretion of the background flow.
For large particles ($\nu\to 0$),
on the other hand, the fourth term, which is the gravitational scattering
by the planet, is dominant and the particles are always repeled away
from the planet.  
For particles with intermediate size
($\nu\sim\Omega_{\rm p}$), all the terms must be considered.  
 Particles located sufficiently far away from the planet are
scattered away because of the contribution from the last term of
equation \eqref{deltab_final}, which is due to the axisymmetric mode of
the perturbed flow structure caused by the planet gravity.  Particles at
the intermediate distance 
($b/H\sim 1$), on the other hand, are attracted towards the planet
because of the third term.  If the particles are close to the planet,
they are again scattered away because of the contribution from
gravitational scattering. 

The equilibrium distance between the
third and fourth term is given by equation \eqref{b_eq_grav}, and this
becomes close to the planet with increasing $\nu$, and it eventually
becomes smaller than the Hill's radius, where we expect that the
particle motion is largely dominated by the planet's gravity and our
treatment of distant encounter breaks down.  Therefore, we conclude that
the particles with intermediate but still smaller drag coefficient, on
one hand, will
accumulate around the planet, but not fully accrete onto the planet.
On the other hand, particles with larger drag coefficient will accrete
onto the 
planet since gravitational scattering is ineffective in this case.
Equating the equilibrium distance given by equation \eqref{b_eq_grav}
to Hill's radius, the critical drag coefficient 
separating these two regimes is of order $10 \Omega_{\rm p}$.

We note that since equation \eqref{deltab_final} is obtained under the
assumption of distant encounter, it does not accurately describe the
dynamics of 
particles located within the Hill's sphere of the planet.
%%%%%%Added in revision%%%%%%%%%%%%%%%%%
In Section \ref{numerical}, we perform a numerical calculation of Hill's
equations and investigate the validity of assumptions in the
derivation of equation \eqref{deltab_final}.
%%%%%%Added in revision end%%%%%%%%%%%%%%%%% 
It needs to be checked whether particles will really
accrete onto the planet when they approach towards the planet.  For
close encounter, the detailed structure of the gas around the planet
must be considered.  Inaba \& Ikoma (2003) have shown that the capture
cross section of particles by the planet may be enhanced when the
atmospheric structure is considered.

\section{Numerical Calculation in Local Coordinate}
\label{numerical}

%\textbf{
%Note: This section is completely rewritten in revision
%}

\subsection{Numerical Methods}

To investigate the accuracy of the description by equation
\eqref{deltab_final} for the evolution of 
orbital element of a particle whose semi-major axis is close to the
planet embedded in a gaseous disk, we  
perform a series of numerical calculations and compare the results with
equation \eqref{deltab_final}.

In our calculation, we solve Hill's equations \eqref{EoM_x} and
\eqref{EoM_y} using fifth-order Runge-Kutta method with variable time
step.  We are primarily interested in the effects of
spiral density wave and the gravitational force by the planet centered
at the coordinate system: we neglect the effect of the global pressure
gradient and mass accretion, $\eta=\zeta=0$.
%%%%%%%%Added in revision%%%%%%%%%%%%%%%%%%
  The velocity field of the gas is therefore
\begin{equation}
 \delta \vect{v}_g = \delta \vect{v}^{(1)} + \delta \vect{v}^{(2)},
\end{equation}
where $\delta \vect{v}^{(1)}$ and $\delta \vect{v}^{(2)}$ are 
first and second order perturbations, respectively, described in section 
\ref{subsec:db_spiral}.  We also 
note that for gravitational force exerted by the planet, we use
\begin{equation}
 \nabla \dfrac{GM_{\rm p}}{\sqrt{x^2+y^2}}.
\end{equation}
We do not use the time-averaged force given by equations \eqref{app_fx}
and \eqref{app_fy}.
%%%%%%%%Added in revision end%%%%%%%%%%%%%%%%%%

%%%%%%%%Added in revision (moved from previous section)%%%%
The modification of gas velocity due to the gravitational force by the
 planet is obtained by second-order perturbation analysis.  
We numerically integrate 
equations \eqref{EoC_hydro_lin}-\eqref{EoMy_hydro_lin} for the first
 order solution and \eqref{EoC_hydro_2nd}-\eqref{EoMy_hydro_2nd} for
 the second order solutions.
We make use of Fourier
transform methods presented by Goodman and Rafikov (2001).  This method
is particularly useful in calculating the gravitational potential
produced by the density fluctuation. 
We fix the spatial resolution
 (therefore, the maximum wave number in Fourier space)
$(\Delta x/H, \Delta y/H)$ to be 
$(9.8 \times 10^{-3}, 7.8 \times 10^{-2} )$, which is enough to resolve
 the effective Lindblad resonance (see e.g., Artymowicz 1993).  We fix
 the box size in the $y$-direction such that $-80H<y<80H$ and impose a
 periodic boundary condition.  For the $x$-direction, the
 resolution in Fourier space in $k_x$ direction is varied from $10^4$ to
 $10^6$ according to $k_y$ in order 
 to resolve very fast oscillation at $k_x \gg k_y$.  We also use the
 linear window function given by Goodman and Rafikov (2001).
 We set softening parameter $\epsilon$ to be $0.01H$.
%%%%%%%%Added in revision (moved from previous section end)%%%%

  The box size of our calculation in $y$-direction is
$-80H<y<80H$, to be consistent with the calculation 
of the modification of the velocity field.
  In integrating Hill's equations, we impose
a periodic boundary condition in the $y$-direction and integrate the
equations of motion until $t=5000\Omega_{\rm p}^{-1}$.
%%%%%%%%%Added in revision%%%%%%%%%%%%%%%%%
By periodic boundary condition, we mean that when the particle has
reached the boundary of the box in $y$-direction, $y=\pm 80H$, it is
reintroduced into the other end of the box, $y=\mp 80H$, while the
values of $x$, $v_x$, and $v_y$ are kept unchanged.
In principle, Hill's equations describe the motion of the particle only
in the vicinity of the planet, and hence there is no reason that the
phase of the epicyclic motion is conserved when it is reintroduced.
However, we 
expect that if the box size is the same as $2\pi r_{\rm p}$, which is
the circumference of the disk at the orbit of the planet, this
periodic boundary treatment may simulate qualitative properties of
the orbit in the global calculations.  The condition
\begin{equation}
 L_y = 2\pi r_{\rm p}
\end{equation}
is realized, in the parameter of our calculations, if we consider a disk
with the aspect ratio $H/r_{\rm p}=2\pi/160 \sim 0.04$.
%%%%%%%%%Added in revision end%%%%%%%%%%%%%%%%%

%%%%%%%%Added in revision%%%%%%%%%%%%%
Once we know the data of $(x,\dot{x},y,\dot{y})$, we can calculate the
osculating elements $(b,h,k)$ by, using equations \eqref{orbit_x} and
\eqref{orbit_y}, 
\begin{equation}
 b = 4x + \dfrac{2}{\Omega_{\rm p}} \dot{y},
  \label{b_instant}
\end{equation}
\begin{equation}
 h = - \dfrac{1}{r_{\rm p} \Omega_{\rm p}} \left[  
	 \dot{x}\sin (\Omega_{\rm p} t) + 2 \left(
	\dot{y}+\dfrac{3}{2}x\Omega_{\rm p} \cos (\Omega_{\rm p} t)
	\right)
 \right],
  \label{h_instant}
\end{equation}
\begin{equation}
 k = \dfrac{1}{r_{\rm p} \Omega_{\rm p}} \left[  
	 \dot{x}\cos (\Omega_{\rm p} t) - 2 \left(
	\dot{y}+\dfrac{3}{2}x\Omega_{\rm p} \sin (\Omega_{\rm p} t)
	\right)
 \right].
  \label{k_instant}
\end{equation}
In the following sections showing results of calculations, we refer to
$(b,h,k)$ using these relations.  These values are not averaged over the
encounter, but are defined at any time of the orbit.
%%%%%%%%Added in revision end%%%%%%%%%%%%%

  The initial condition
of the calculation is that the particles are located at the edge of the
box and moving in a circular
motion.  When a particle
approaches to the planet and the mutual distance becomes smaller than
half of the Hill's radius of the planet, we stop the calculation since
we are primarily interested in the distant encounter.  Such particles
should be trapped by the planet's gravity and eventually accrete onto
the planet.  In order to obtain a smooth results in the runs with large
drag coefficient, we make use of smoothing methods 
described in Appendix \ref{app:sphint} in calculating the perturbed gas
velocity at the location of the particle from the data obtained by
hydrodynamic equations.
We fix the
background density to be $10^{-10} \mathrm{g/cm}^3$, which is an order of
magnitude smaller than the Minimum Mass Solar Nebula (Hayashi et
al. 1985).  Toomre's $Q$ parameter of the disk at $1 \mathrm{AU}$ with
is
\begin{equation}
 Q^{-1} \sim \dfrac{G\rho}{\Omega_{\rm p}^2} \sim 10^{-4}
\end{equation}
  We assume the mass of the planet as 
$GM_{\rm p}/Hc^2=10^{-2}$ that corresponds to $0.2M_{\oplus}$ for
$c=10^5\mathrm{cm/s}$ and $H=0.05\mathrm{AU}$.
In the following sections, we show the results without the gravitational
force by gas.  We have checked that gas gravity does not affect the
results in our parameter range.  We also neglect the effect
of global pressure gradient and steady mass accretion in our
calculation, $\eta=\zeta=0$, in order to see the effect of spiral wave 
and planet's gravity.

\subsection{Results}
\label{subsec:numericalresults}

\subsubsection{Properties of Orbital Evolution}

\begin{figure}
 \plotone{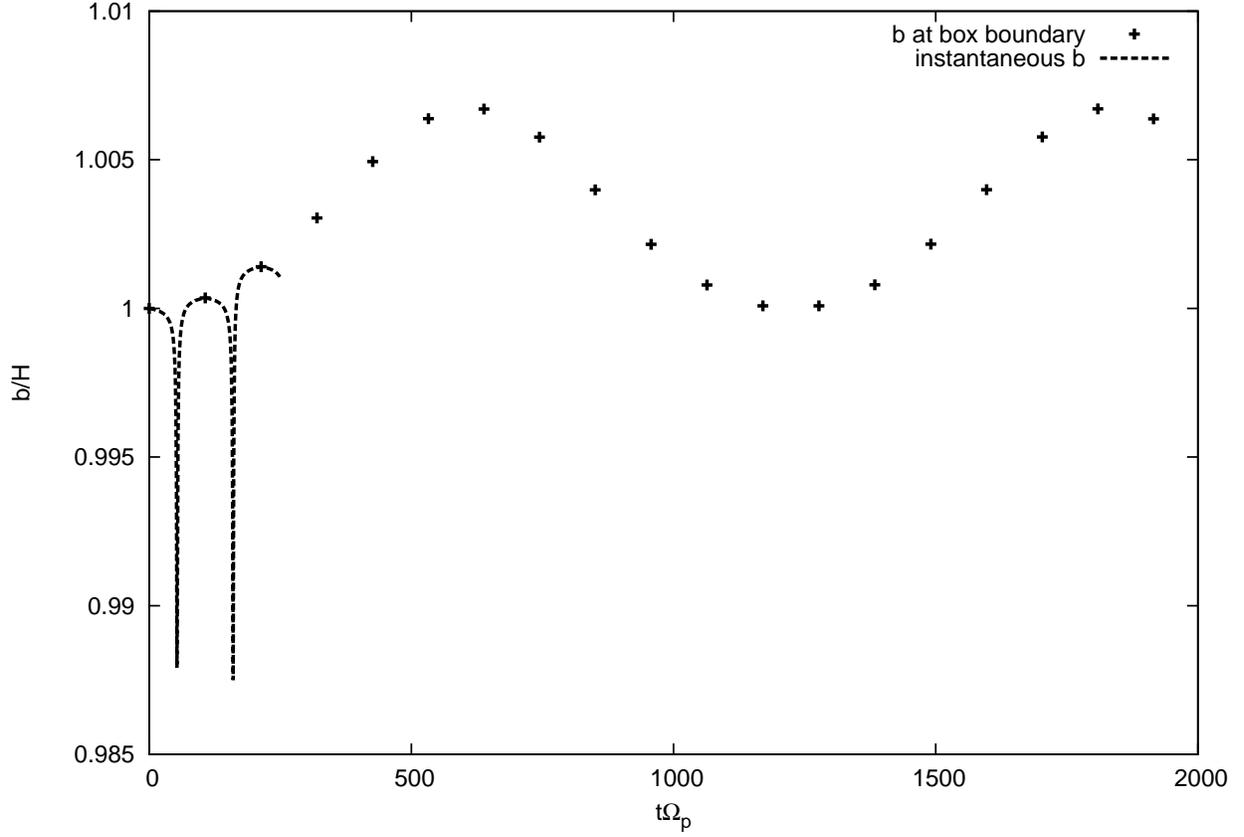}
 \caption{Evolution of $b$ for a particle with zero drag coefficient
 initially located at $b=H$.  Symbols show the values of $b$
 obtained at the box boundary.  The
 value of $b$ during the encounter is also indicated by dashed line upto
 $t=250\Omega_{\rm p}^{-1}$.}
 \label{fig:threebody_b1}
\end{figure}

%%%%%%%%%%Added in revision%%%%%%%%%%%%
We first review the properties of orbital evolution of a particle with
zero friction force, which is obtained by setting $\nu=0$.  Figure
\ref{fig:threebody_b1} shows the evolution of $b$ for a particle
initially located at $b=H$.  The values of $b$ are obtained according to
equations \eqref{b_instant} when the particle is at the box boundary.
Also indicated in Figure \ref{fig:threebody_b1} is the variation of
$b$ for $0<t<250\Omega^{-1}$.  The values of $b$ varies rapidly when the
particle encounters with the planet, when $y\sim0$.  We note that since
the box size is $160H$, time taken for the particle to cross the box is
$\sim 160/(3/2) \sim 100$ Kepler times. 

\begin{figure}
 \plotone{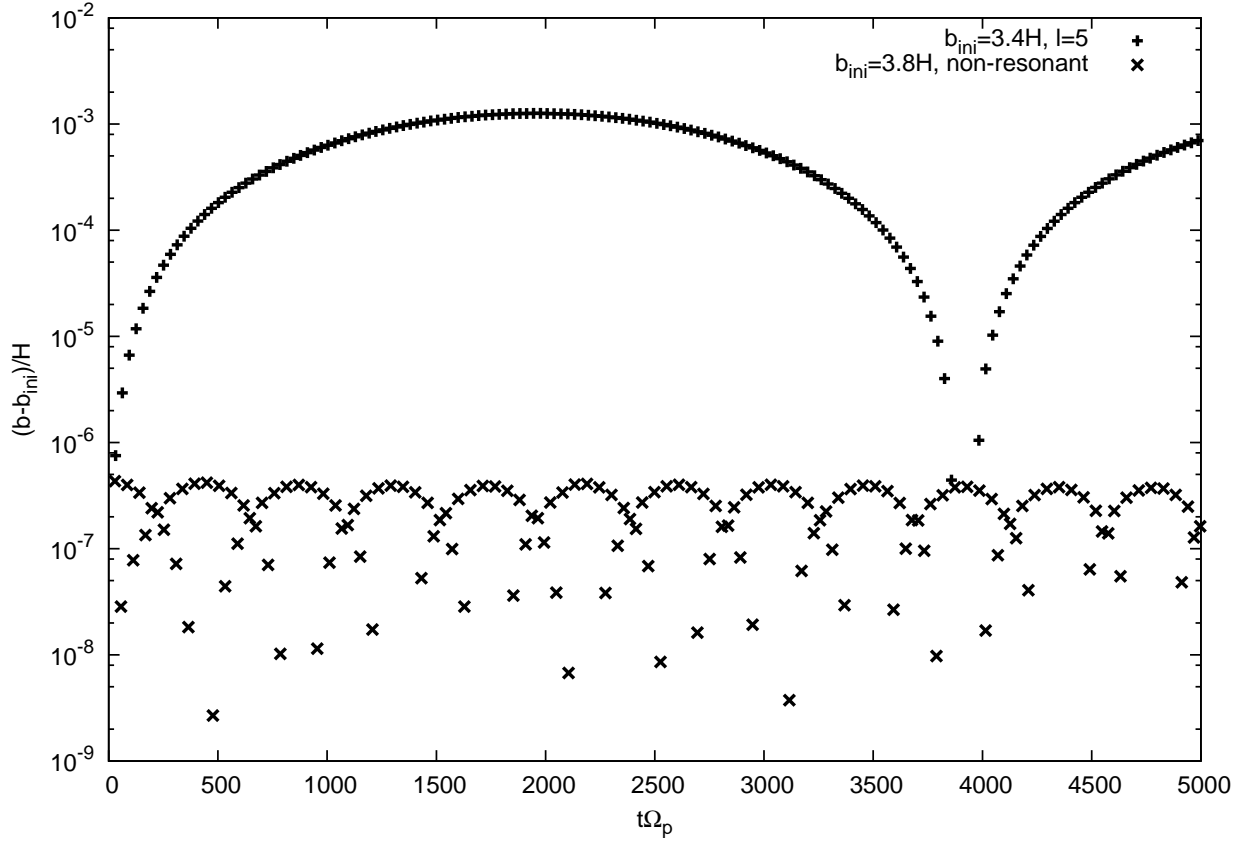}
 \caption{Evolution of $b$ for a particle with zero drag coefficient
 initially located at $b_{\rm ini}=3.4H$ (plus) that corresponds
 to $l=5$ case in equation \eqref{resonance_cond} and 
 $b_{\rm ini}=3.8H$ (cross) that does not satisfy equation
 \eqref{resonance_cond}.  Vertical axis indicates the value of
 $(b(t)-b_{\rm ini})/H$.  The value of $b(t)$ is obtained when the
 particle reaches the box boundary.  Note that vertical axis of this
 figure is in logarithmic scale and the values smaller than $10^{-9}$
 are not plotted.  Horizontal axis shows time.}
 \label{fig:threebody_res}
\end{figure}

It is indicated that the semi-major axis difference $b$ first increases
but then shows an oscillation with the period $\sim 1200$ Kepler times.
This oscillation is caused by the excitation of eccentricity by the
perturbation due to the planet.  If the particle comes into the box with 
finite eccentricity, the distance between the particle and the planet at
the closest approach is different depending on the phase of the
epicyclic motion, and the semi-major axis of the particle can increase
or decrease.  The period and amplitude of this oscillation depend on the
location of the particle.  If the phase of the epicyclic motion among
successive encounters between the 
planet and the particle matches, the period of the oscillation becomes
longer and the amplitude becomes larger.  The condition for strong
amplification is given by
\begin{equation}
 \Omega_{\rm p}T=2\pi l, 
  \label{resonance_cond}
\end{equation} 
where $l$ is an integer and $T$ is the time taken to cross the box in
$y$-direction 
\begin{equation}
 T = \dfrac{L_y}{(3/2)b\Omega_{\rm p}}.
\end{equation}
Figure \ref{fig:threebody_res} compares the evolution of semi-major axis
of the particle starting at $l=5$ position ($b=3.4H$) and at
$b=3.8H$.  It is clear that the particle at $l=5$ position experiences
the long-term, strong oscillation of semi-major axis.

Equation \eqref{resonance_cond} indicates that there is a resonance
between 
the box crossing time and the epicyclic motion of the particle, and this
appears as a result of our treatment of periodic boundary in Hill's
coordinate.
  Thus, in general, we should not consider this resonance to be physical
  if we arbitrary choose the box size in the y-direction.  
However, if we take the box size $L_y$ as the circumference of
the disk, $2\pi r_{\rm p}$, the crossing time $T$ corresponds
to the period of synodic encounter.  Therefore, it may be possible to 
interpret this resonance as $j/(j+1)$ resonances in global
problems.  In section \ref{global}, we explore the
correspondence between the Hill's approximation and the full global
problem. 
%%%%%%%%%%Added in revision end%%%%%%%%%%%%

\begin{figure}
 \plotone{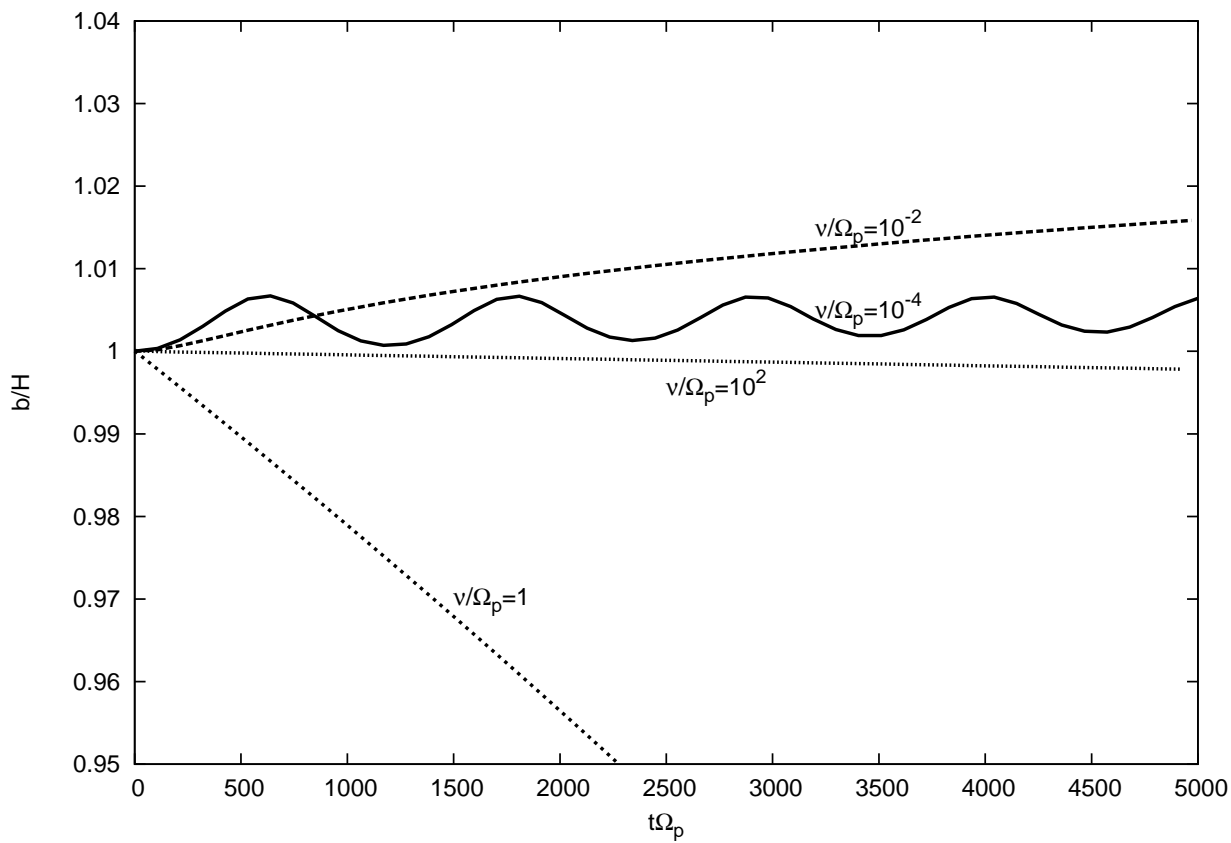}
 \caption{Evolution of $b$ for particles initially located at $x=H$ with
 different drag 
 coefficient.  Planet mass corresponds to $GM_{\rm p}/Hc^2=10^{-2}$.
Particles with small drag coefficients shows oscillation,
 while particles with large drag coefficient shows a systematic decrease
 or increase of semi-major axis.}
 \label{fig:bevol_nu}
\end{figure}

%%%%%%%%%Modified in revision%%%%%%%%%%%
We now consider the problem with gas drag.  
Figure \ref{fig:bevol_nu} shows the evolution of $b$ for particles with
various drag coefficients initially located at $b/H=1$ as a function of
time.  The values of $b$ when the particles reach at the edge of the box
are plotted.
  Particles with drag coefficient larger than 
$\nu/\Omega_{\rm p}=10^{-2}$ shows a systematic change of semi-major
axis while that with $\nu/\Omega_{\rm p}=10^{-4}$ shows a systematic
change plus oscillation.  In a simple restricted three-body problem,
where drag coefficient $\nu$ equals zero, semi-major
axis does not show 
a systematic change but only oscillation, as discussed in previous
paragraphs (see also Figures \ref{fig:threebody_b1} and
\ref{fig:threebody_res}).  The gas drag 
causes a systematic change of Jacobi energy 
and thereby results in a systematic change of particle's angular
momentum.  The timescale during which the oscillation of semi-major
axis damps is given by $\sim \nu^{-1}$.
Since the oscillation of the semi-major axis comes from the excitation
of eccentricity during the synodic encounter, this oscillation does not
appear when the eccentricity is damped during successive encounters.  
If the eccentricity is damped during successive encounters, the
assumption of initially circular orbit in deriving equation
\eqref{deltab_final} is valid for every encounter.  Therefore, although
we have computed only one encounter, it is possible to use equation
\eqref{deltab_final} to predict the
long-term evolution of the orbital semi-major axis 
especially for small bodies with large drag coefficient.
%%%%%%%%%Modified in revision end%%%%%%%%%%%

\subsubsection{Comparison between Analytic and Numerical Calculations}

%%%%%%%%%Added in revision%%%%%%%%%%%%
We now show to what extent equation \eqref{deltab_final} describes 
the evolution of the orbit of a particle encountering with a planet
embedded in a gaseous nebula.  Since the particles show a systematic
change of semi-major axis, we take the average radial velocity of the
particle by calculating
\begin{equation}
 \dfrac{b(t_{\rm fin}) - b(t_{\rm ini})}{t_{\rm calc}},
  \label{average_db_calc}
\end{equation}
where $t_{\rm ini}=0$ is the initial time,
$t_{\rm fin}$ is the last time when the particle reaches
the box boundary before the calculation is stopped at
$t=5000\Omega^{-1}$, and $t_{\rm calc}=t_{\rm fin}-t_{\rm ini}$.  The
precise value of $t_{\rm fin}$ is obtained by numerical calculation, and
it varies with the value of $b$.
Semi-major axis of the particle at each time is calculated by equation
\eqref{b_instant}.
%%%%%%%%%Added in revision end%%%%%%%%%%%%

\begin{figure}
 \plotone{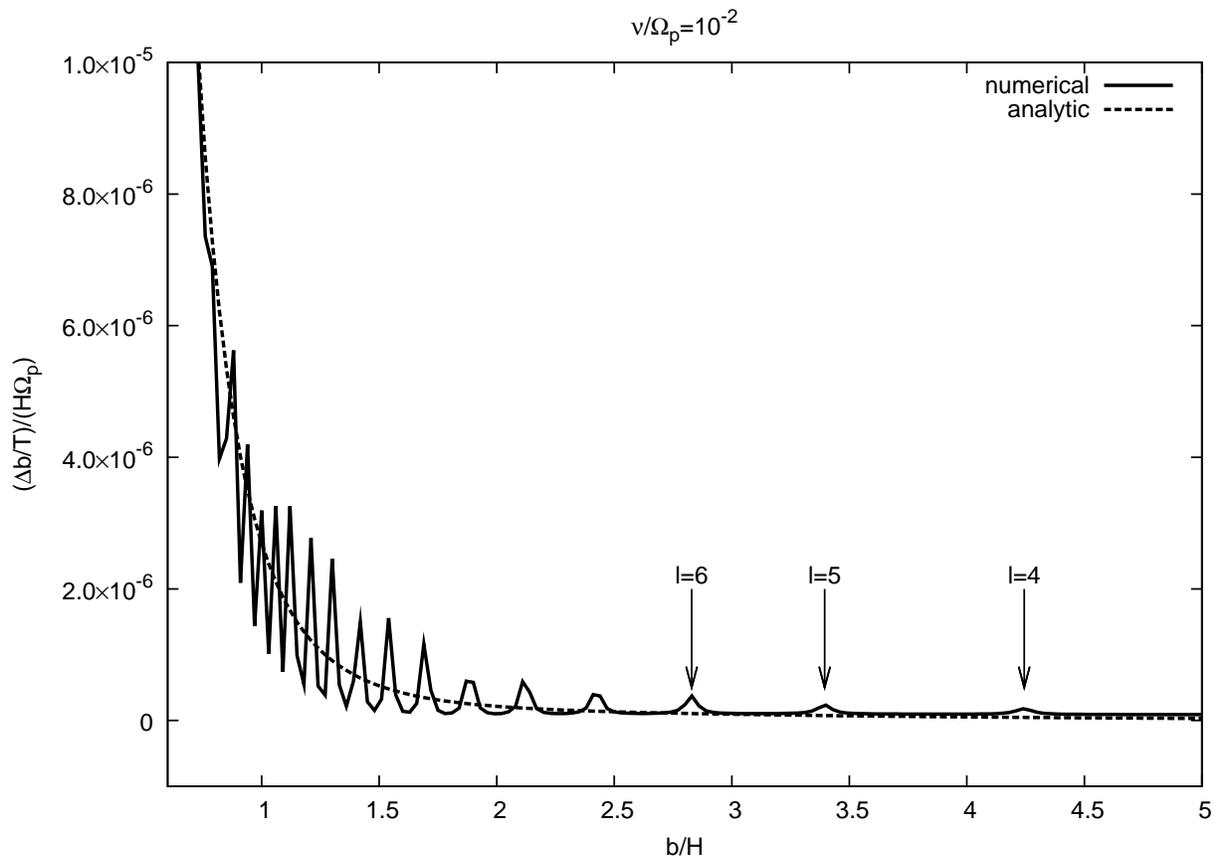}
 \caption{Comparison between the local numerical calculation (solid
 line) and analytical formula (dashed line) of $\Delta b / T$ for
 particles with $\nu/\Omega_{\rm p}=10^{-2}$.  
 The value of $\Delta b/T$ is the average of a number of encounters and
 calculated using equation \eqref{average_db_calc}.
 The gravitational force by the planet and the
 spiral pattern of the gas around the planet are both considered.}
 \label{fig:bdecay_Hill_d10m}
\end{figure}

\begin{figure}
 \plotone{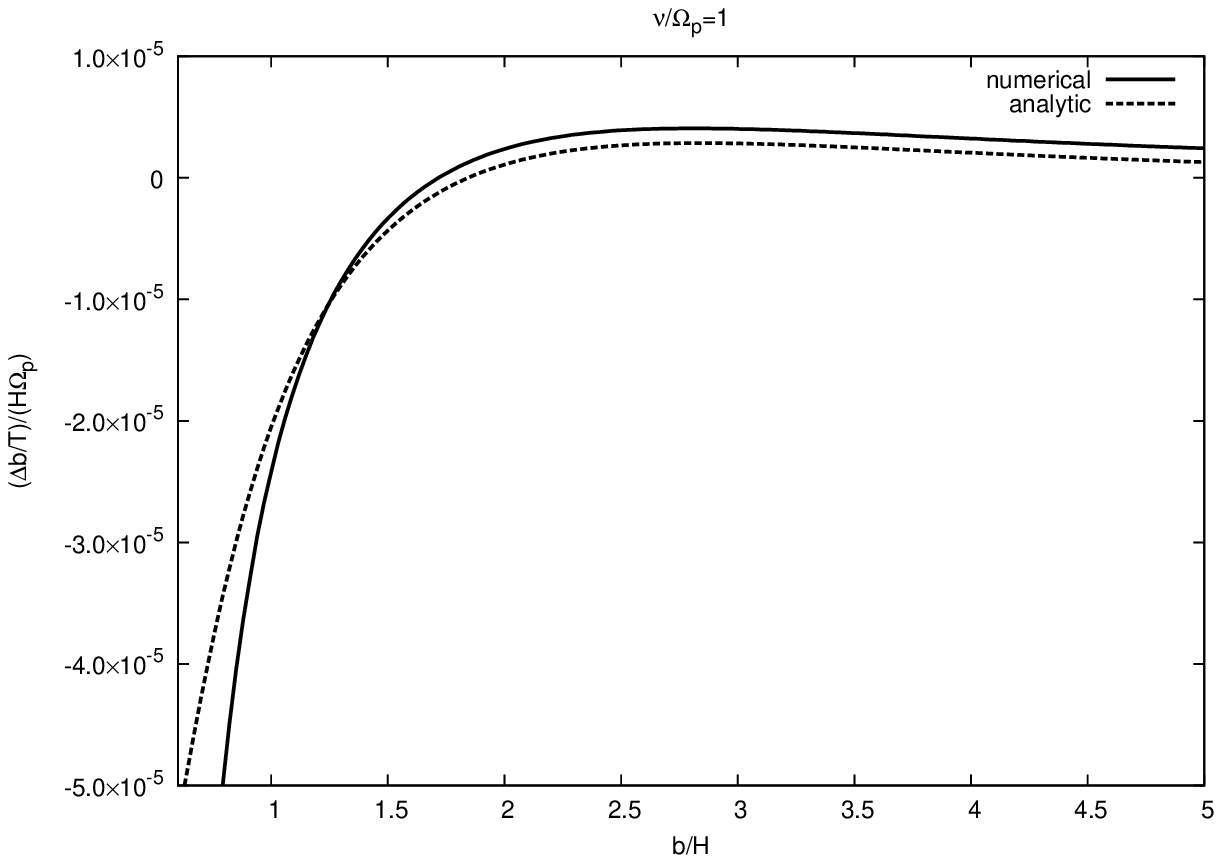}
 \caption{Same as Figure \ref{fig:bdecay_Hill_d10m} but for
 particles with $\nu/\Omega_{\rm p}=1$.}
 \label{fig:bdecay_Hill_d1m}
\end{figure}

\begin{figure}
 \plotone{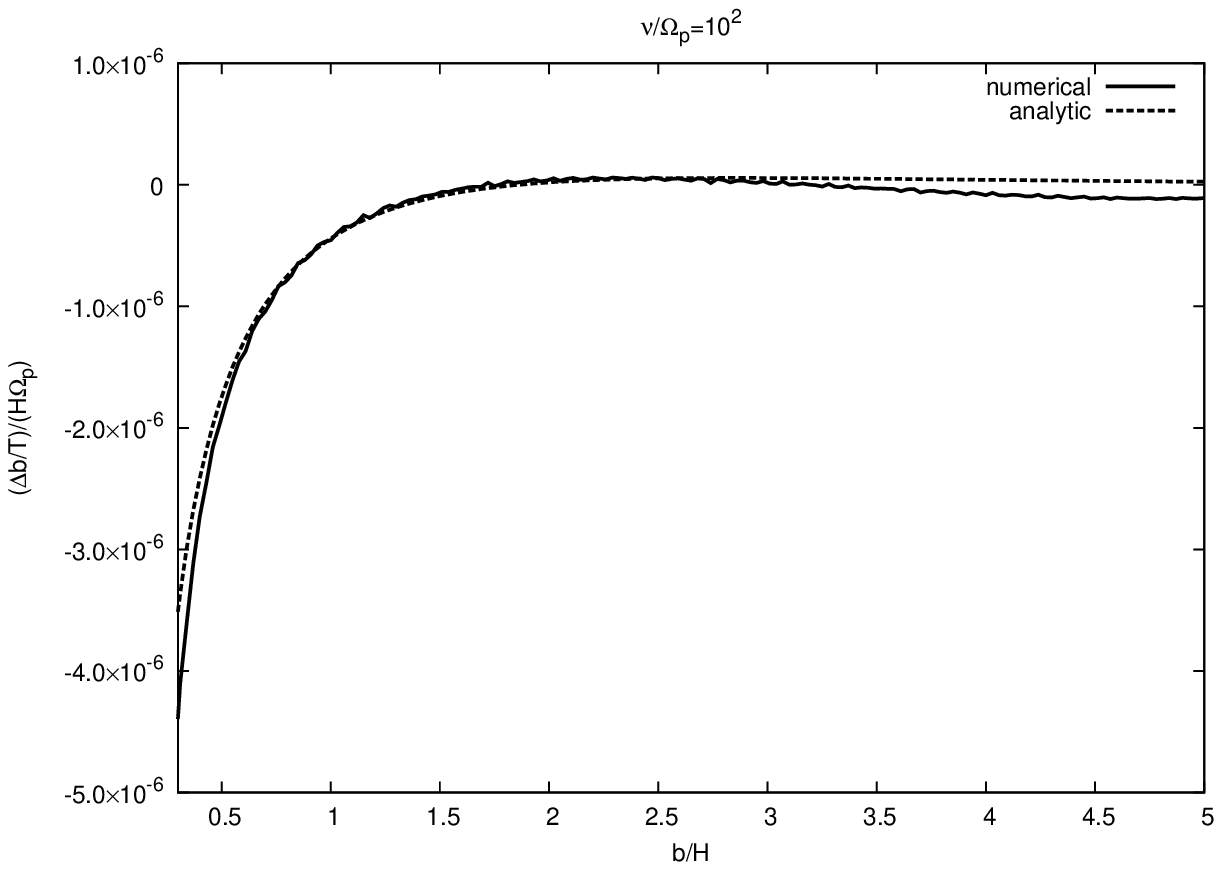}
 \caption{Same as Figure \ref{fig:bdecay_Hill_d10m} but for
 particles with $\nu/\Omega_{\rm p}=10^2$.}
 \label{fig:bdecay_Hill_d10cm}
\end{figure}

Figures \ref{fig:bdecay_Hill_d10m}-\ref{fig:bdecay_Hill_d10cm} show
the results of long-term evolution of semi-major axis of particles
initially located at various position with 
$\nu/\Omega_{\rm p}=10^{-2}$, $1$, and $10^2$ respectively.  Horizontal
axes of Figures \ref{fig:bdecay_Hill_d10m}-\ref{fig:bdecay_Hill_d10cm}
are the initial semi-major axis of the particle, $b(t_{\rm ini})$, and
the vertical axes show the average rate of change of semi-major axes
obtained by equation \eqref{average_db_calc}.
  We also
plot our analytic formula given by \eqref{deltab_final}.
 It is clear
that the analytic formula \eqref{deltab_final} describes the results of
numerical calculations well, at least quantitatively, for all values 
of drag coefficient.  Deviation from the analytic value for
$\nu/\Omega_{\rm p}=1$ at $b/H \gtrsim 2$ comes from the deviation of
the data of $\overline{\delta v_y^{(1)}}$ from the analytic value shown
in Figure \ref{fig:dvy_asym_profile} while for small $b$, we expect that
the numerical results deviate from the analytic calculation since
eccentricity effect is not negligible.
  The match between numerical calculation and 
analytic formula is better for particles with large drag coefficient,
since their eccentricities remain smaller.   We think that the
deviation from the analytic value for $\nu/\Omega_{\rm p}=10^2$ at
$b/H\gtrsim 3$ is due to smoothing. 

One significant difference between analytic results and numerical
calculation is the spike-like structures that appear in the calculations
with $\nu/\Omega_{\rm p}=10^{-2}$.
This structure is due to the resonance effect given by equation
\eqref{resonance_cond}.  
 We show in Figure
\ref{fig:bdecay_Hill_d10m} the location of resonance with $l=4$, $5$,
and $6$.  It is clear that $\Delta b/T$ shows a spike at this location.
We have not found such resonance structure for calculations with 
$\nu/\Omega_{\rm p} \gtrsim 1$.  This is because eccentricity is
sufficiently damped between successive encounters.
   
\begin{figure}
 \plotone{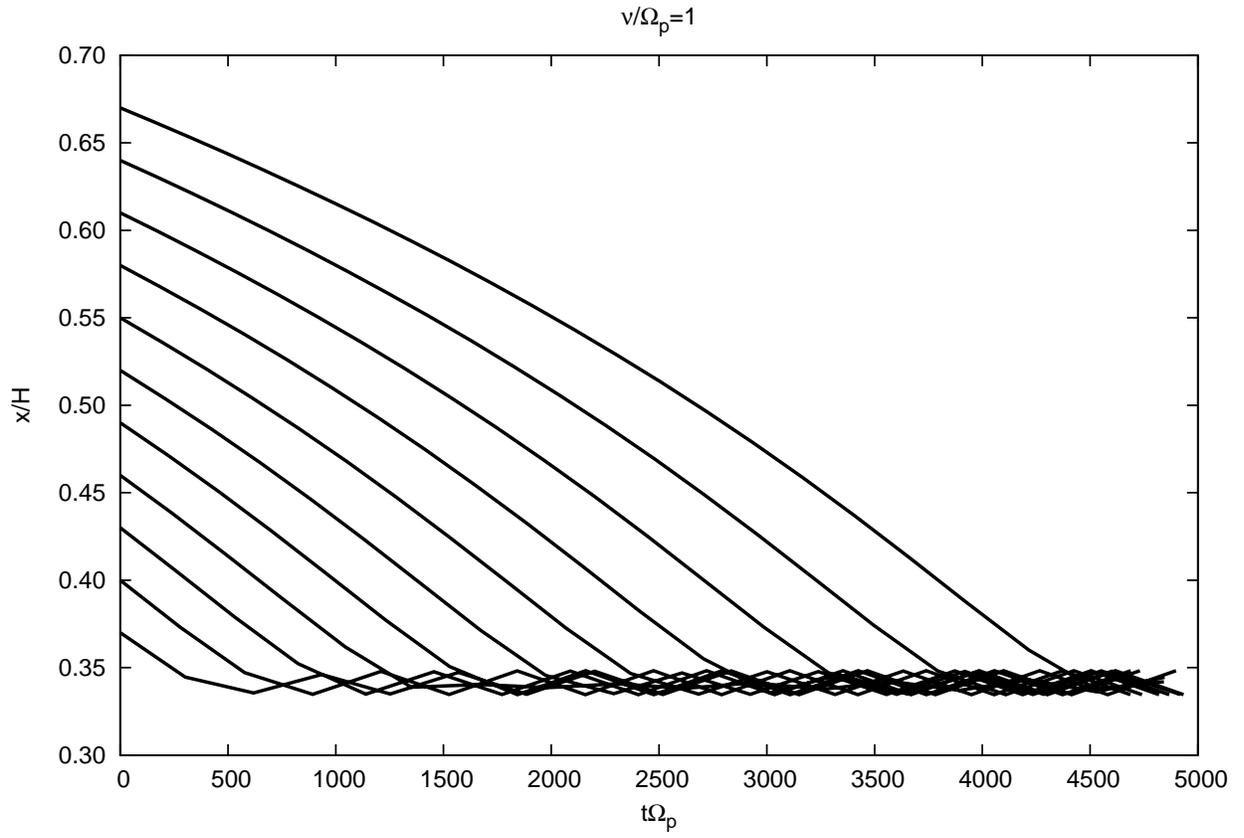}
 \caption{Long-term evolution of the semi-major axis of the particles
 with $\nu=\Omega_{\rm p}$.  The mass of the planet corresponds to  
$GM_{\rm p}/Hc^2=10^{-2}$.  The calculations are done for the disk model
 with zero pressure gradient and zero accretion flow, $\eta=\zeta=0$.}
 \label{fig:long_nu1}
\end{figure}

%%%%%%%%Modified in revision%%%%%%%%%%
Another interesting result is the motion of the particle with 
$\Omega_{\rm p}\sim \nu$.
 Figure \ref{fig:long_nu1}
shows the evolution of the semi-major axes of the particles with 
$\nu/\Omega_{\rm p}=1$.  It is clear that the particles migrate
towards the planet and stagnate at $b\sim 0.35H$.
Since we set $\eta=\zeta=0$ in our calculation, the orbital change of
the particle is caused by the third, fourth, and fifth terms of equation 
\eqref{deltab_final}.  The last term is negligible for small $b$
compared to the third term.  
  Equation \eqref{deltab_final} indicates that
these particles accumulate near the planet's orbit where
gravitational 
scattering represented by the fourth term of equation
\eqref{deltab_final} balances with the attraction towards the planet
represented by the third term, which is the result of combination of the 
planet's gravity and the gas drag, as
discussed in Section \ref{subsec:db_planet}. 
  The condition for the balance between these two effects is given by
  equation \eqref{b_eq_grav}.  For chosen parameters in numerical
  calculations, $r_{\rm H}=0.15H$ so the equilibrium distance is given
  by $\sim 0.3 H$, which is close to the value obtained in numerical
  calculation $\sim 0.35H$. 
 We consider that the small difference comes from the effect of
  eccentricity and close encounter.
  This effect seems analogous to ``the shepherding effect''.  
We note that this effect
appears when the first term of equation
\eqref{deltab_final}, $\Delta b$ by global
pressure gradient, is small compared to the third term.  This condition
is expressed in terms of $\eta$:
\begin{equation}
 |\eta| < \dfrac{r_{\rm p}}{b} \dfrac{r_{\rm H}^3}{r_{\rm p}^3} 
  \sim \dfrac{r_{\rm p}}{b} \dfrac{M_{\rm p}}{M_{\ast}} .
\end{equation}
  If $\eta$ is of the order of square of disk aspect ratio, as in
  standard parameters, this condition is rewritten as
\begin{equation}
 \dfrac{b}{r_{\rm H}} \lesssim 
  \left( \dfrac{M_{\rm p}}{M_{\ast}} \right)^{2/3}
  \left( \dfrac{r_{\rm p}}{H} \right)^2.
\end{equation}
For $M_{\rm p}/M_{\ast}=10^{-6}$ and $H/r_{\rm p}=0.05$, this gives
$b/r_{\rm H}\lesssim 0.2$, which is too close to the planet and the
assumption of distant encounter is violated.  Kary et al. (1993) has
shown that, in the presence of global pressure gradient, a particle with 
$\nu \sim \Omega_{\rm p}$ falls onto the 
central star so rapidly that it bypasses the planet without being
trapped.
%%%%%%%%Modified in revision end%%%%%%%%%%
 
%%%%%%%%%section added in revision%%%%%%%%%
\subsection{Limitation of Analytic Formula}
\label{subsec:limitation_local}

%\textbf{
%Note: This section is added in revision
%}

Equation \eqref{deltab_final} is derived by assuming that the
deflection angle of the particle by each encounter with the planet is
small (distant encounter) and that the particle returns to a circular
orbit before the next encounter.  

First, the assumption of initially circular orbit breaks down if we
consider the particle with small friction.  The oscillation of
semi-major axis observed in figure \ref{fig:bevol_nu} for particles with
$\nu=10^{-4}\Omega_{\rm p}$ indicates that the damping of eccentricity
between successive encounters is insignificant for these particles and
that the finite eccentricity effect should be taken into account.
Quantitatively, the period of synodic encounter must be longer than the
particle's stopping time.  Therefore, our treatment is limited in the
cases with
\begin{equation}
 \dfrac{2\pi r_{\rm p}}{(3/2) b \Omega_{\rm p}} \gtrsim \dfrac{1}{\nu}, 
\end{equation}
where the left hand side is the synodic period and the right hand side
is stopping time.  Rearranging this condition, we have
\begin{equation}
 \dfrac{b}{H} \lesssim \dfrac{4\pi}{3} \dfrac{r_{\rm p}}{H}
  \dfrac{\nu}{\Omega_{\rm p}}.
\end{equation}
Adopting standard parameters, we have
\begin{equation}
 \dfrac{b}{H} \lesssim 80 
  \left( \dfrac{0.05}{H/r_{\rm p}} \right)^{-1}
  \left( \dfrac{\nu/\Omega_{\rm p}}{1} \right) .
  \label{condition_e}
\end{equation}

\begin{figure}
 \plotone{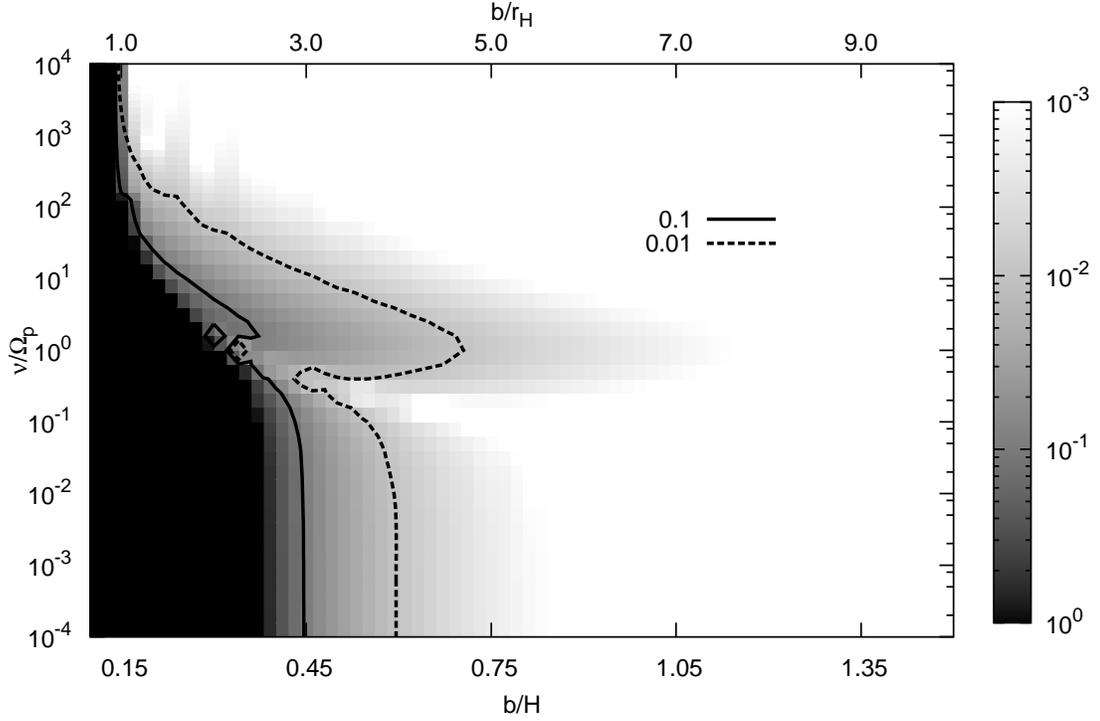}
 \caption{Deflection of the particle's semi-major axis after the first
 encounter with the planet.  The amount of deflection is quantified by
 equation \eqref{deflection_rate}.  Horizontal axis denotes
 the particle's initial semi-major axis and vertical axis denotes the
 particle's drag, or equivalently, size.  Initial semi-major axes are
 shown in terms of scale height (bottom axis) and Hill's radius (top
 axis).  Gray scale shows the amount of deflection, and the contours of
 10\% (solid line) and 1\% (dashed line) deflection are shown.
 Particles in black regions are trapped by the planet (that is, mutual
 distance between the particle and the planet becomes smaller than half
 of Hill's radius) or in horseshoe orbit.}
 \label{fig:size_deflection}
\end{figure}

We now consider the assumption of distant encounter.  If the particle's 
semi-major axis is too close to the planet, the assumption of small
deflection angle, or distant encounter, breaks down.
In Figure \ref{fig:size_deflection}, we plot the amount of deflection
of the particle's semi-major axis after the very first encounter with
the planet.  We quantify the amount of deflection by
\begin{equation}
 \dfrac{|b(t_1) - b(t_2)|}{|b(t_1)|},
  \label{deflection_rate}
\end{equation}
where $t_1=t_{\rm ini}=0$ is the initial time, and
$t_2 \sim L_y/(3/2)b\Omega_{\rm p}$ is the time when the particle
reaches the box boundary after the first encounter.  The precise value
of $t_2$ is obtained by numerical calculations.  Particles
in black regions are trapped by the planet, that is, in our criterion,
mutual distance between the planet and the particle becomes smaller than
half of the Hill's radius, or in horseshoe orbit.  Note the similarity
of the shape of the contour of 1\% deflection with the contour shown in
Figure \ref{fig:db_contour}.

\begin{figure}
 \plotone{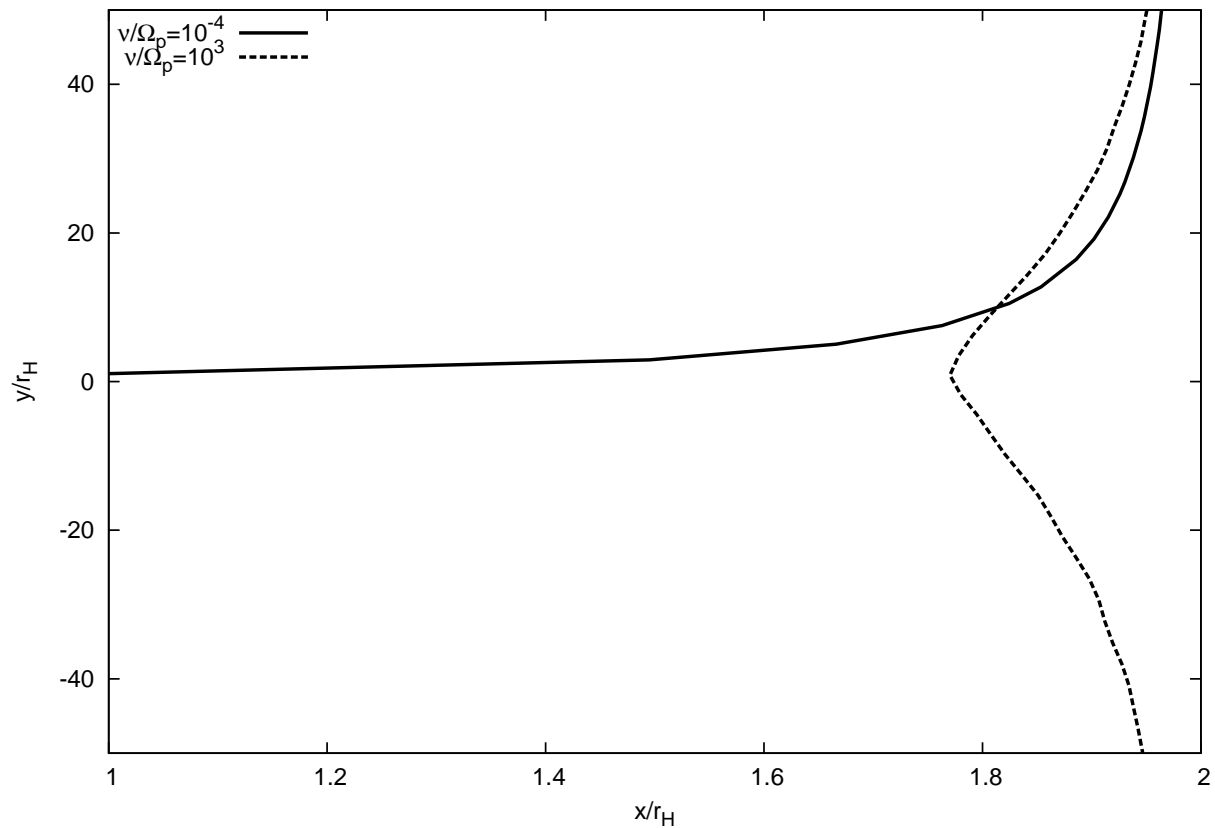}
 \caption{Examples of the orbit of the particle when they encounter with
 the planet.  Particles with $\nu/\Omega_{\rm p}=10^{-4}$ (solid line)
 and $\nu / \Omega_{\rm p}=10^3$ (dashed line) are shown.  Both
 particles are initially located at $b=2r_{\rm H}$.  Only the region
 $(1r_{\rm H}<x<2r_{\rm H},-50r_{\rm H}<y<50r_{\rm H})$ is shown.  The
 particle with small drag is strongly perturbed by the planet and
 eventually captured within the Hill's radius, while that with large
 drag can escape owing to the drag by the background gas.} 
 \label{fig:orbit_2hill}
\end{figure}

In three-body problem with zero drag force, it is known that the
assumption of small encounter is violated for particles with 
$b\lesssim 2.5 r_{\rm H}$ if we consider the very first encounter.  Our
calculation is 
consistent with this for particles with small drag, 
$\nu \ll \Omega_{\rm p}$.  However, for particles with large drag,
$\nu\gg\Omega_{\rm p}$, it can be seen that this condition is relaxed
and it is indicated that the assumption of small deflection angle can be
used even down to $b \sim r_{\rm H}$.  Figure \ref{fig:orbit_2hill}
shows the orbits of particles initially located at $b=2r_{\rm H}$.
Orbits of particles with $\nu/\Omega_{\rm p}=10^{-4}$ and $10^3$ are
shown.  Particles with very large drag coefficient can remain in the
orbit close to the initial semi-major axis since gas drag prevents it
from falling onto the planet.

After multiple encounters, the condition of distant encounter is
violated for particles with $b\lesssim2.5-2\sqrt{3}r_{\rm H}$ in
three-body problem without gas drag.  The contour of 1\% deflection in
Figure \ref{fig:size_deflection} passes close to $2\sqrt{3}r_{\rm H}$ in
the limit of small drag.  Therefore, we infer that this line gives the
limitation of our analytic treatment assuming small deflection.  This
assumption is better for particles in wider range of semi-major axes for
larger drag.

%%%%%%%%%section added in revision end%%%%%%%%%

%%%%%section rearranged in revision%%%%%
\section{Comparison between Local and Global Calculation}
\label{global}

%\textbf{
%Note: This section is completely rewritten in revision
%}

%%%%%%%This section is completely rewritten in revision%%%%%%
So far we have performed analytic and numerical calculations of the
orbits of particles using local coordinate systems.  It remains to be
checked how our local treatment simulates the real orbit in global
coordinate.  

We have performed a global calculation with the gas rotating at
Keplerian rotation velocity where we have neglected 
the effect of the spiral pattern around the planet, global pressure
gradient, and gas accretion towards the central star, $\eta=\zeta=0$.
  We solved the equations of 
motion in a frame rotating with the planet and integrate until
the particle encounters with the planet for 100 times.  We calculated
the evolution of the semi-major 
axis of a particle at various initial semi-major axes and obtain
 $\Delta b/T$ over the calculation time. 
 In the
following, we show calculations in which the particle is located at the
opposition point initially, but we have checked that the initial
locations do not affect the results as long as we consider particles
with $\nu/\Omega_{\rm p} \geq 10^{-2}$.
Particles that reach the distance with the planet less than Hill's
radius are removed.  We set disk aspect ratio $H/r_{\rm p}=0.04$, which
gives $2\pi r_{\rm p}=L_y$, where $L_y=160H$ is the box size we have
used in calculations with Hill's equation.  For planet mass, we assumed 
$M_{\rm p}/M_{\ast}=6\times 10^{-7}$, which corresponds to 
$GM_{\rm p}/Hc^2=10^{-2}$, the value we have adopted in calculations
with Hill's equations.  In order to compare the results of global
calculations with Hill's equations, we also integrate Hill's equations
\eqref{EoM_x} and \eqref{EoM_y} by setting $\delta \vect{v}_{\rm g}=0$.  

\begin{figure}
 \plotone{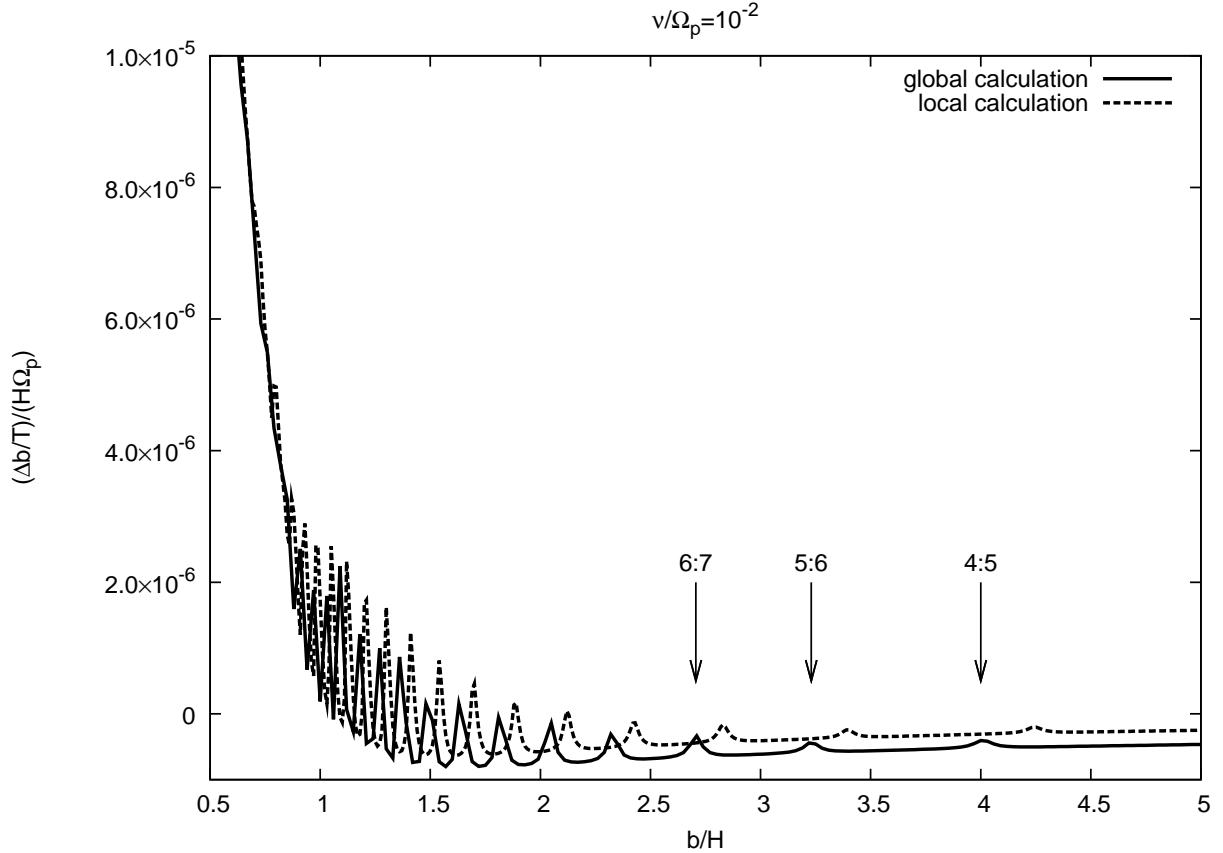}
 \caption{Comparison between the global numerical calculation (solid
 line) and local calculation (dashed line) of
 $\Delta b / T$ for 
 particles with $\nu/\Omega_{\rm p}=10^{-2}$.  The average of the change
 of the semi-major axis of the particle after 100
 encounters with the planet is shown.
  Horizontal axis shows the initial location of the particle normalized
 by disk scale height, $b/H=(r-r_{\rm p})/H$.
 We use disk aspect ratio $H/r_{\rm p}=0.04$ and the mass ratio between
 the planet and the central star is $M_{\rm p}/M_{\ast}=6\times10^{-7}$.
 The disk aspect ratio is chosen in such a way that the value of $L_y$
 we have adopted in the integration of Hill's equation corresponds to
 $2\pi r_{\rm p}$, and the planet mass is chosen in such a way that
 $GM_{\rm p}/Hc^2=10^{-2}$, which is also the value we have used in
 local calculations.  Keplerian rotation of the gas is assumed so
 $\eta=\zeta=0$ and there is no modification of the gas structure by the
 planet's gravity.  Note that global calculations and local calculations
 are in good agreement except for the location of the resonances.  The
 position of $4:5$, $5:6$, and $6:7$ resonances are indicated by arrows. 
}
 \label{fig:globalHill_d10m}
\end{figure}

\begin{figure}
 \plotone{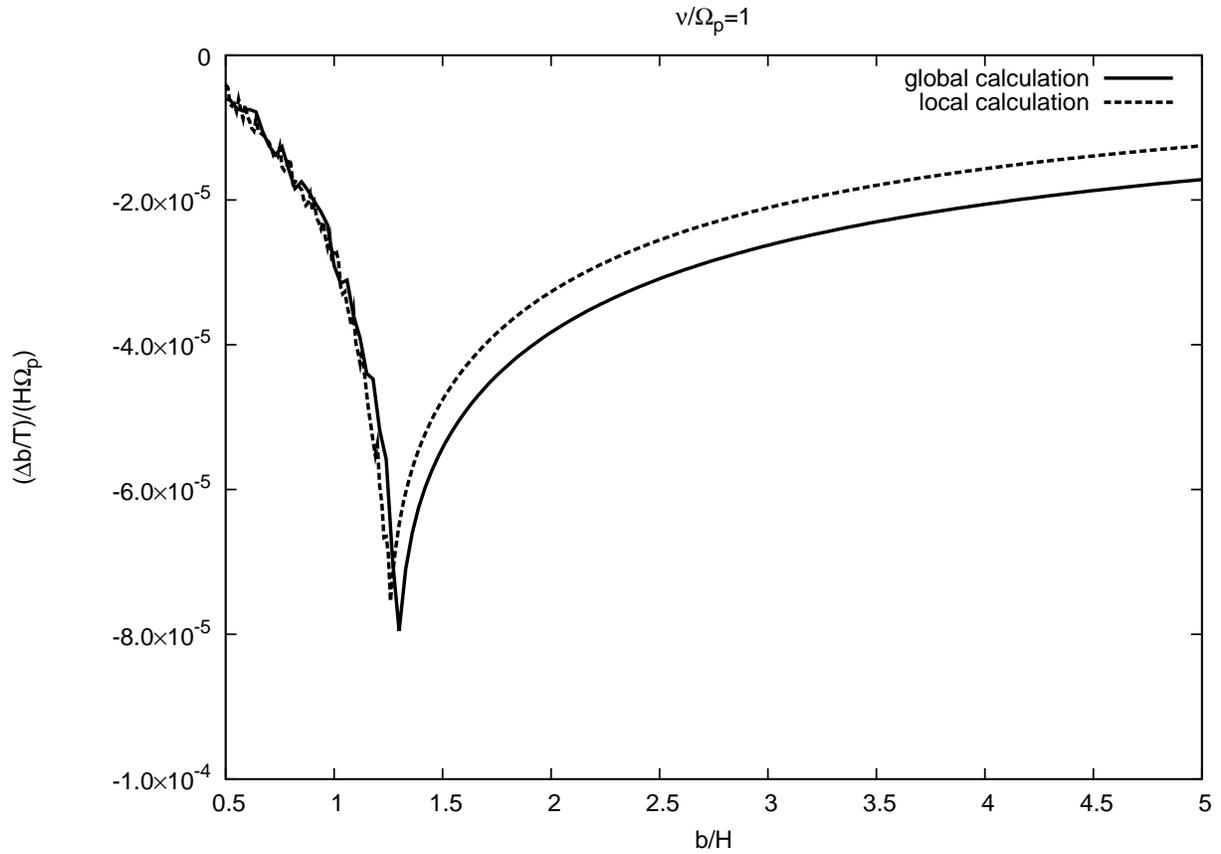}
 \caption{Same as Figure \ref{fig:globalHill_d10m} but for
 particles with $\nu/\Omega_{\rm p}=1$.  Particles initially located at
 $b/H < 1.4$ has been trapped in the vicinity of the resonance so the
 average rate of orbital change over 100 encounters is small.  See also
 Figures \ref{fig:long_nu1} and \ref{fig:long_nu1_global} for the time
 evolution of semi-major axis difference $b$ 
 for particles with $b/H \lesssim 1.4$.}
 \label{fig:globalHill_d1m}
\end{figure}

\begin{figure}
 \plotone{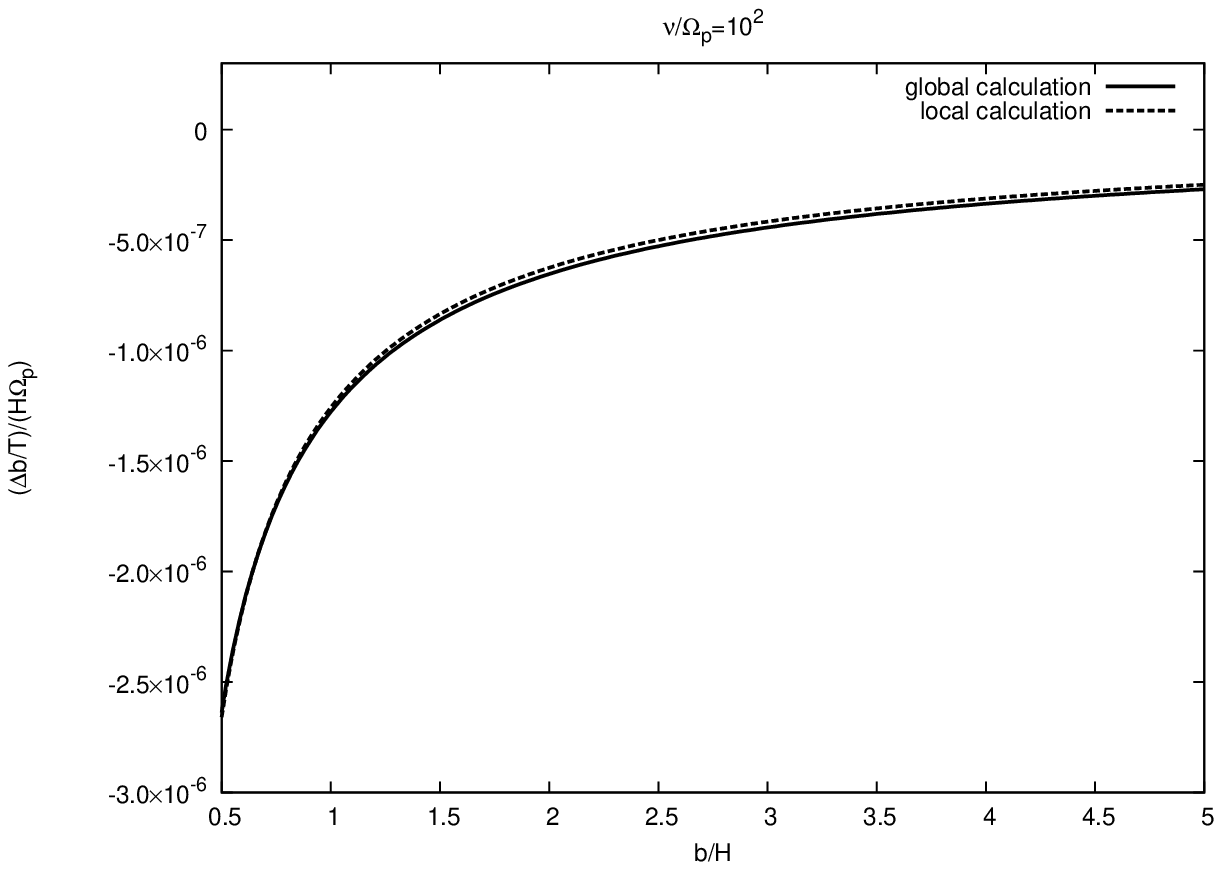}
 \caption{Same as Figure \ref{fig:globalHill_d10m} but for
 particles with $\nu/\Omega_{\rm p}=10^{2}$.}
 \label{fig:globalHill_d10cm}
\end{figure}

Figures \ref{fig:globalHill_d10m}-\ref{fig:globalHill_d10cm} compare the
results of $\Delta b/T$ obtained by global and local calculations.
Horizontal axes of Figures
\ref{fig:globalHill_d10m}-\ref{fig:globalHill_d10cm} show the initial
semi-major axes of the particle and the vertical axes show the average
rate of change calculated by equation \eqref{average_db_calc}.  In
global calculations, $b(t_{\rm fin})$ is the semi-major axis difference
between the particle and the particle calculated when
the particle has reached the opposition point after 100 encounters.

  For all parameters, global and local calculations show a reasonably
  good agreement.  Relative difference between global and local runs
is most significant for particles with 
${\nu}/{\Omega_{\rm p}}=1$.  We infer that this is because in the case
of ${\nu}/{\Omega_{\rm p}}=1$, the radial motion of the particle is the 
fastest, and the effect of curvature is most effective.  In runs with
${\nu}/{\Omega}_{\rm p}=10^{2}$ and ${\nu}/{\Omega}_{\rm p}=10^{-2}$,
radial motion of the particle is smaller by one or two orders of
magnitude than calculations with ${\nu}/{\Omega}_{\rm p}=1$.

In runs with $\nu/\Omega_{\rm p}=10^{-2}$, we
observe a disagreement in resonance locations, since curvature effects
are not taken into account.  However, since the disagreement is
less than 10\%, it may be still possible to say, for example, that $l=4$ 
resonance in Hill's coordinate mimics $4:5$ resonance in global
coordinate.  Note that the relative strength of the resonance matches
well with global runs.

Global analysis of $j/(j+1)$ resonances including drag force is done by 
Greenberg (1978).  It
is shown that eccentricity and the angle between the longitude of
conjunction and the longitude of pericenter of the particle converges to
a fixed value  with the timescale of $\nu^{-1}$, but depends on the
particle's semi-major axis.  The value of eccentricity is maximum at the
resonance.  We have observed similar behavior in local calculations.
It is actually possible to do analytic calculation similar to Greenberg 
(1978) in Hill's system and derive orbital evolution including the
effect of resonance.  We do not show it here since it is necessary to
do all the analyses presented in Section \ref{analytic} again including
the effect of finite box size.  We simply note here that this resonance
effect does not appear in our formulation presented in Section
\ref{analytic} since we have neglected the effect of finite length in
the $y$-direction in deriving equation \eqref{deltab_final}.

\begin{figure}
 \plotone{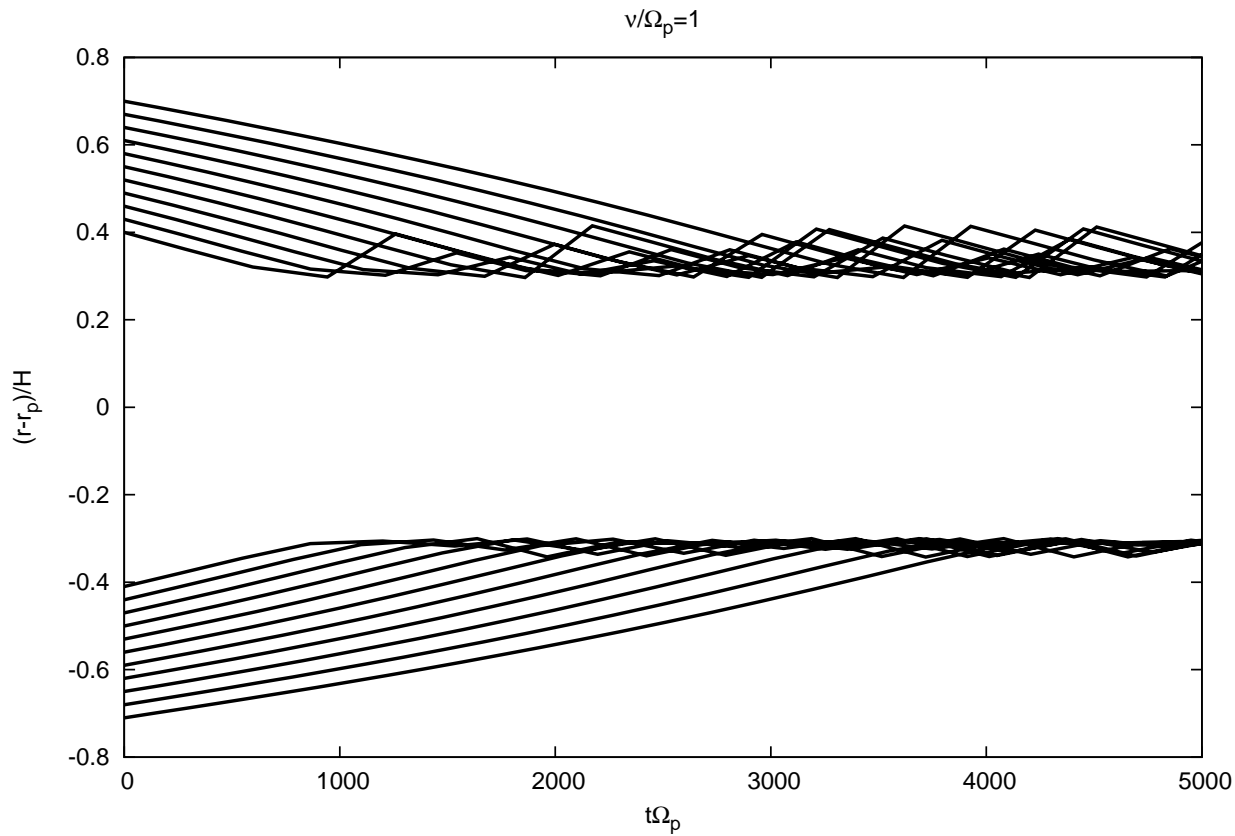}
 \caption{Evolution of semi-major axis of particles with
 $\nu/\Omega_{\rm p}=1$ in global run with $\eta=\zeta=0$ and the
 gas is assumed to rotate at Kepler velocity.  Osculating elements of
 the particle at the opposition point is plotted.  Particles in both
 interior and exterior orbits migrate towards the planet and trapped in
 the orbit of $\sim 0.3H$.   This figure corresponds to Figure
 \ref{fig:long_nu1} that shows the results of local calculations.}
 \label{fig:long_nu1_global}
\end{figure}

In runs with $\nu/\Omega_{\rm p}=1$, particles initially at
 $b\lesssim 1.4H$ shows the decrease of the average rate of semi-major
 axis decay.  This is because the particle is trapped in the orbit in
 the vicinity of the planet as a result of shepherding effect discussed
  for calculations with Hill's
 equation in Section \ref{subsec:numericalresults}.  Figure
 \ref{fig:long_nu1_global} shows the evolution of the 
 semi-major axis of the particle.  Osculating elements at the opposition 
 points are plotted.  Comparing with Figure \ref{fig:long_nu1}, it is
 clear that the shepherding effect discussed in local calculations also
 exists in global runs.  We note that the amplitude of oscillation after
 particles are trapped is larger in global runs than local calculations,
 and the location of the shepherding orbit is slightly closer to the
 planet than the local calculations.  Note in passing that in
 equation (4.11) of Adachi et al. (1976), it is indicated that for
 $\eta=0$, damping of eccentricity causes a decrease in semi-major axis.
 However, Adachi et al. (1976) do not take into account
 the effects of the planet's gravity.  The shepherding presented here is
 entirely caused by the presence of the planet.

By comparing global calculations with local runs, it is indicated that
as long as the orbits close enough to the planet are considered, local
calculations reproduce the results of global calculations well even if
curvature effect is neglected.  However, the effect of constant box size
 in the $y$-direction causes the small difference between the
global and local runs.  This is indicated by the mismatch between the
locations of the resonances.  It is also indicated that the local
calculations reproduce the results of global calculations most
efficiently for small particles,
 such as $\nu/\Omega_{\rm p} \lesssim 10^{-1} - 10^{-2}$. 
 Local approach that is used in Section \ref{analytic} is valid when
 semi-major axis difference $b$ is much smaller than the scale of the
 semi-major axis of the planet:
\begin{equation}
 \dfrac{b}{r_{\rm p}} \ll 1,
\end{equation}
which is equivalent to
\begin{equation}
 \dfrac{b}{H} \ll \dfrac{r_{\rm p}}{H} .
  \label{condition_local}
\end{equation}
Note that in the usage of equation \eqref{deltab_final} there is another
limitation of close encounter, which has already been addressed in
Section \ref{subsec:limitation_local}.
%%%%%%%This subsection is completely rewritten in revision%%%%%%

%%%%%%%section rearranged in revision%%%%%%
\section{Discussion: Dust Gap Opening}
\label{dustgap}

\subsection{Order-of-Magnitude Estimate of Dust Gap Opening Criterion}

Paardekooper and Mellema (2004, 2006) showed that dust gap may be opened
up even
if the planet mass is too small to open up a gap in a gas disk using
numerical calculation.  We reconsider this result from the analytical
point of view in this section.

Since there is no long-term radial motion of fluid elements, it is
essential in dust gap formation that gas and dust moves in a different
velocity.  Terms that are important in equation \eqref{deltab_final} are
therefore those proportional to 
$\nu\Omega_{\rm p}/(\nu^2 + \Omega_{\rm p}^2)$ for small particles.  

We first consider the disk without global pressure gradient.  
In this case, the motion of small dust particles are mainly dominated by
the third term of equation \eqref{deltab_final} when they are as close
to the planet as $b\lesssim2H$ and by the last term when $b$ is larger.
The third term is the attraction of particles by the planet's gravity
and they migrate towards the planet.  The last term on the other hand 
originates from the axisymmetric flow structure
around the planet.  This term causes dust particles migrate away from
the planet.  In either case, dust particles around the planet with
$b \sim H$ forms a gap around the planet.  Taking $b\sim H$, very rough
order-of-magnitude estimate (neglecting numerical coefficients of order
unity) gives the migration timescale of dust particles as
\begin{equation}
 \tau_{\rm gap} \equiv b \dfrac{T}{\Delta b} \sim 10^{4-5} T_{\rm K}
  \left(\dfrac{10^{-5}}{M_{\rm p}/M_{\rm c}}  \right)
  \left( \dfrac{H/r_{\rm p}}{0.05} \right)^2
  \left( \dfrac{(\nu^2+\Omega_{\rm p}^2)/\nu \Omega_{\rm p}}{10^2}
  \right) 
\end{equation}  
where $T_{\rm K}$ denotes the Kepler timescale and $M_{\rm c}$ is the
mass of the central star.  This is the timescale
of dust particles changing its orbit by the order of scale height and
gives the timescale of the formation of dust gap with width of the
order of the scale height.

If there is global pressure gradient, the first term of equation
\eqref{deltab_final} may dominate over the
terms considered above.  Since global pressure gradient causes
systematic inward or outward motion of particles, dust gap around the
planet is not expected to form if this term dominates.
The timescale of orbital change of dust
particles by the order of scale height due to the first term of equation
\eqref{deltab_final} is
\begin{equation}
 \tau_{\eta} \sim 10^{3-4} T_{\rm K}
  \left( \dfrac{10^{-3}}{|\eta|} \right) 
  \left( \dfrac{H/r_{\rm p}}{0.05} \right)
  \left( \dfrac{(\nu^2+\Omega_{\rm p}^2)/\nu \Omega_{\rm p}}{10^2}
  \right) .
\end{equation}
If the magnitude of $\eta$ is of order $(H/r_{\rm p})^2 \sim 10^{-3}$,
which is a 
standard value, dust gap may not appear around the planet since 
$\tau_{\eta}<\tau_{\rm gap}$.
However if $\eta$ is as small as $10^{-5}$, dust gap formation may be
possible.  The value of $\eta$ depends on disk model and can take values
of wide range.  From theoretical point of view, small value of $\eta$ is
preferable in order for meter-size particles to survive in a Minimum
Mass Solar Nebula.  One mechanism of decreasing $\eta$ around ice line
is suggested by Kretke and Lin (2007).

Comparing the first and the third term of equation \eqref{deltab_final},
we obtain the criterion of dust gap formation.  Again, for $b\sim H$,
and neglecting numerical coefficients of order unity, we have, for the
gap formation criterion,
\begin{equation}
 |\eta| \dfrac{M_{\rm c}}{M_{\rm p}} \dfrac{H}{r_{\rm p}} \lesssim 1 .
\end{equation}
We note that this criterion does not depend on drag coefficient of the
particles.  However, the implicit assumption is that gravitational
scattering, which is the fourth term of equation \eqref{deltab_final},
is ineffective, which is proportianl to 
$\Omega_{\rm p}^2/(\nu^2+\Omega_{\rm p}^2)$.  Therefore, this condition
applies to particles with $\nu\gtrsim \Omega_{\rm p}$.
Taking $\eta \sim (H/r_{\rm p})^2$ and $H/r_{\rm p} \sim 0.05$, this
gives $M_{\rm p}/M_{\rm c} \gtrsim 10^{-4}$, which explains the
condition obtained by Paardekooper and Mellema (2004, 2006).

Timescale constraint of dust gap formation may be obtained by comparing
the above obtained timescale with the timescale of planet migration or
gas dispersal timescale.  Timescale of the type I planet migration of
the radial distance comparable to the scale height is given by
Tanaka et al. (2002) as 
\begin{equation}
 \tau_{\rm mig I} \sim \Omega_{\rm p} \dfrac{M_{\rm c}}{M_{\rm p}}
  \dfrac{M_{\rm c}}{\Sigma r_{\rm p}^2} 
  \left( \dfrac{H}{r_{\rm p}} \right)^3 .
\end{equation}
Therefore, comparing this with $\tau_{\rm gap}$, we have
\begin{equation}
 \dfrac{r_{\rm p}}{H}
 \dfrac{\Sigma r_{\rm p}^2}{M_{\rm c}}
 \dfrac{\nu^2+\Omega_{\rm p}^2}{\nu \Omega_{\rm p}} \lesssim 1 .
\end{equation}
Dust gap formation is efficient at late stage of planet formation when
gas starts to dissipate and type I migration timescale becomes long.
If $H/r_{\rm p}=0.05$ and $\Sigma r_{\rm p}^2/M_{\rm c}=10^{-3}$, this
criterion gives $\nu/\Omega_{\rm p} \lesssim 50$.  
If there is no type I migration, on the other hand,  this
constraint is relaxed and dust gap formation constraint is given by the
timescale of disk gas dispersal, $\tau_{\rm gas}$.  If $\tau_{\rm gas}$
is of the order of $10^6$ years, gap of particles with 
$\nu/\Omega_{\rm p} \lesssim 10^4$ may be opened up.

\subsection{Model of Radial Distribution of Dust
   Particles}
\label{discuss:dustdistribution}

In order to investigate the qualitative behavior of dust particles
around the planet in long time, we model the motion of particles in the
disk by one-dimensional advection equation
\begin{equation}
 \dfrac{\partial N(t,b)}{\partial t} + \dfrac{\partial}{\partial b} 
  \left[ v_b (b) N(t,b) \right] = 0
  \label{dustadvection_eqn}
\end{equation} 
where $v_b (b)$ is the radial velocity of the dust particles whose
semi-major axis difference with the planet is $b$
and $N(t,b)$ is the appropriately normalized number of the particles
with semi-major axis difference $b$ at time $t$.  We use equation
\eqref{deltab_final} for $v_b$ and solve \eqref{dustadvection_eqn} for
$10^6 \mathrm{yr}$ for particles with various drag coefficients.  We
have used second-order scheme with monotonicity
condition (van Leer 1977).  We fix $H$ and $c$ for simplicity in our
computational domain and a planet with $GM_{\rm p}/Hc^2=10^{-1}$ is
fixed at the origin.
   Planet mass corresponds to $2M_{\oplus}$ for
$H=0.05\mathrm{AU}$ and $c=10^5\mathrm{cm/s}$.
The computational domain is $r_{H}<b<10H$ 
and dust particles are homogeneously distributed initially:
$N_0\equiv N(t=0,x)=1$ in this region.  We have assumed that there is no
dust particle out of this domain.  For $T$ that appears in equation
\eqref{deltab_final}, we have assumed
\begin{equation}
 T = 2\pi / |\Omega_K(r_p+b) - \Omega_K(r_p)|
\end{equation} 
where $\Omega_K$ is Keplerian angular velocity and $r_p$ is the location
of the planet. 

\begin{figure}
 \plotone{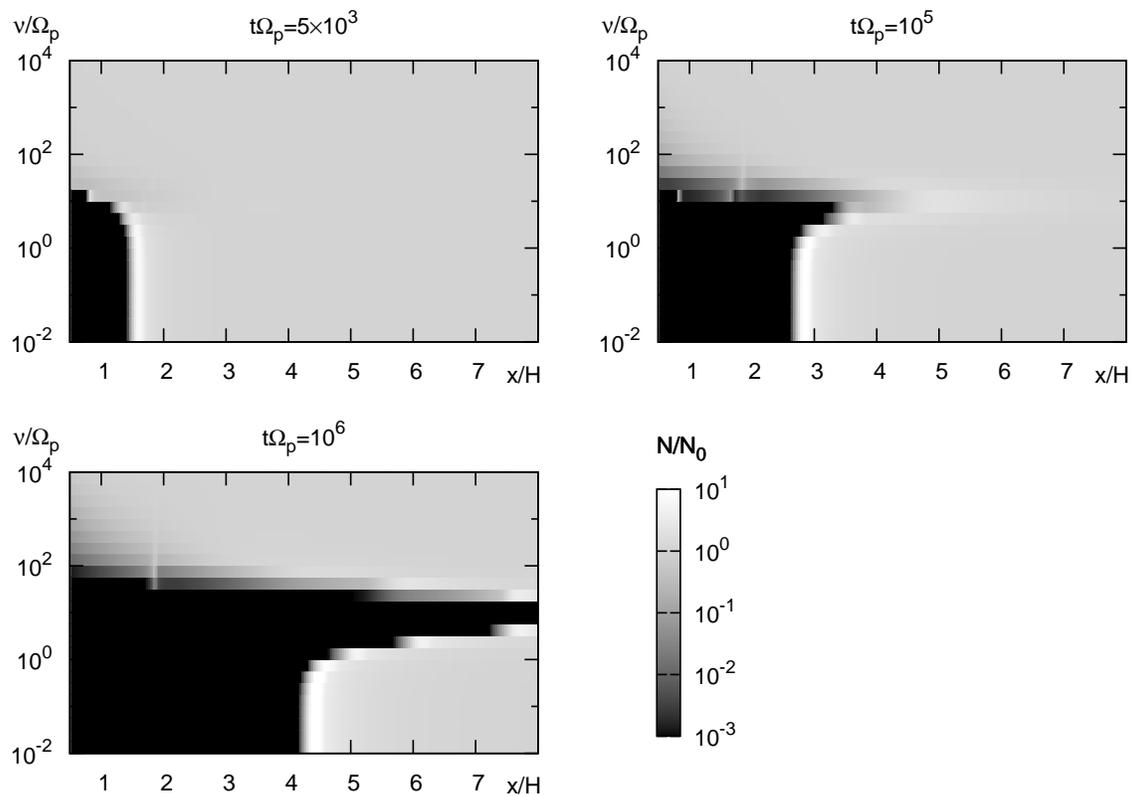}
 \epsscale{0.95}
 \caption{Time evolution of the distribution of dust particles around
 $2M_{\oplus}$ embedded
 in a solar nebula at $t\Omega_{\rm p}=5\times 10^3$ (top left),
 $t\Omega_{\rm p}=10^5$ (top right), and
 $t\Omega_{\rm p}=10^6$ (bottom left) with various $\nu$.
 The horizontal axis denotes the semi-major axis difference between
 the planet and the particle.  Gray scale indicates the fraction of
 the remaining particles.  The
 pressure gradient factor $\eta$ is set to be zero.}
 \label{fig:advection}
\end{figure}

Figures \ref{fig:advection} shows the snapshots of the distribution of
dust particles of various drag coefficients at 
$t\Omega_{\rm p}=5\times10^{3}$, $t\Omega_{\rm p}=10^{5}$, and  
$t\Omega_{\rm p}=10^{6}$ for $\eta=0$.
  The gap of particles with $\nu \sim 10 \Omega_{\rm p}$
has been opened up after several $10^5$
Kepler time.  The gap of particles $\nu\lesssim \Omega_{\rm p}$
 may be the artefact
of boundary condition, since the radial velocity of particles at the
nearest edge of the planet is positive but we have assumed there is no
dust particles outside the calculation domain.  However, there is no
boundary effect for small particles with large drag coefficient, since
radial velocity of dust particle is negative at the inner edge and
positive at the outer edge.  We note that since it takes about $10^5$
years for dust gap formation for the planet 
with $2M_{\oplus}$, this effect is small
 if the gas pressure gradient is significantly large, $\eta>0$.
  However,
for larger mass planets, planet perturbation on the dust motion is large
and dust gap formation is possible even in the presence of pressure
gradient, as seen in the order-of-magnitude estimate provided in the
previous section.  

Finally, we briefly discuss the observational implication of dust gap
in the disk with zero pressure gradient. 
The width of the dust gap that forms around a low mass planet is 
several times the scale height.  
Therefore, it might be challenging to detect
a low mass planet embedded in a gas disk at $1\mathrm{AU}$.  However, if
there is a planet at several tens of AU, the width of
the gap can be of the order of several AU, which can be resolved in 
near future.  The gap formation timescale
in this case is of order $10^{5-6}$ years, 
which might be comparable to disk
life time.  Therefore, it may be possible to find a low mass planet by
observing a dust gap in a gaseous nebula, even if the planet itself is
too small to be observable.

\section{Summary and Future Prospects} \label{summary}

In this paper, we have investigated the motion of particles embedded in
a gas disk in the presence of a small mass planet.  This is the 
full analytic calculation of the particle motion including the gas disk
and the planet considering the non-axisymmetric pattern of the gas around
the planet produced by the gravitational interaction between the planet
and the gas.

Our main result is equation \eqref{deltab_final} which
describes the change of the particle semi-major axis, which has been in
circular motion initially, after one distant encounter with the planet.
In addition to well-known particle migration towards the central star
due to the pressure gradient (the first term) and the gravitational
scattering by the planet (fourth term), 
we have derived (1) the effect of steady mass accretion (the second term), 
(2) attraction of particles towards the planet due to the planet's
gravity (the third term), and (3) contribution of the  
structure of the gas disk produced by the planet's gravity (the fifth
term).
We find that only the axisymmetric structure of the gas
contributes to the semi-major axis evolution of particles and the spiral
structure is ineffective in the absence of vorticity source.

All the terms considered, in the absence of the global
pressure gradient and steady accretion flow towards the central star, we
find that (see Figure \ref{fig:db_contour})
 (1) particles with
 $\nu \gtrsim \Omega_{\rm p}$ will migrate towards the 
 planet if they are close to the planet because of the gravitational
 attraction by the planet, but they migrate away from the planet
 if they are far away, primarily due to the effect of the gas
 structure,
 (2) large particles with 
$\nu \ll \Omega_{\rm p}$ will be scattered away from the planet because
of the gravitational scattering, and
(3) there is a parameter range where particles are scattered away from
the planet when they are close to the planet while attracted towards the
planet if they are located at far away, thereby particles are
accumulated at an equilibrium distance from the planet (i.e.,
shepherding by gas drag and gravitational scattering). 
%%%%%%%Added in revision%%%%%%%%
 This shepherding
mechanism acts for particles in both interior and exterior orbits as
long as the radial pressure gradient of the gas is negligible.
%%%%%%%Added in revision%%%%%%%%

%%%%%%%modified in revision%%%%%%
We have checked the validity of our formula by solving three-body
problem with and without Hill's approximations.  Our main assumptions in
deriving equation 
\eqref{deltab_final} are (1) initially circular orbit, (2) distant
encounter, and (3) local approximation. 
Validity of the first assumption is justified when the eccentricity of
the particle is damped within the time taken for each synodic encounter
that is given by equation \eqref{condition_e}.  The valid region for the
second 
assumption is shown by the contour of small deflection in Figure
\ref{fig:size_deflection}.  The condition for the third assumption is
simply given by equation \eqref{condition_local}.  
Hill's approximation is good 
as long as we consider the orbits in the vicinity of the planet. 
In short, our formula is especially useful in predicting the motions of
dust particles with stopping time $t_e \Omega_{\rm p} \lesssim 10^{2}$.
For particles with smaller drag, our treatment predicts only
quantitative behavior of non-resonant orbital evolution of particles
whose semi-major axis is close to that of the planet.
We also note that since we have made use of linear calculation for the
flow, our treatment is 
restricted to low mass planets with the mass up to the orders of several
tens of Earth mass.
%%%%%%%modified in revision end%%%%%

Using equation \eqref{deltab_final}, we have discussed the criterion for
dust gap formation.  The condition depends on the value of global
pressure gradient, mass of the planet, and disk scale height.  It is
possible to derive the condition suggested by Paardekooper and Mellema
(2004, 2006), and we have derived more general criteria.
We have calculated the qualitative
long-term behavior of dust particles around a planet with
$2M_{\oplus}$.
It is indicated that the gap of small dust particles with width of the
order of several scale height may be opened up
after several $10^5$ Kepler time. This long timescale has not been
reached by full numerical simulations yet.

We have simplified the description of the drag force by assuming the
force proportional to relative velocity with the gas.  For large particles,
however, the drag force is proportional to the square of the relative
velocity, which is not treated in our calculation.  We have also made a
simplification by considering two-dimensional problem.  In considering
the vertical motion of particles, it is necessary to do
three-dimensional calculations.  The effects of magnetic field, 
self gravity, and turbulence of the disk will be the subjects in this
line of analytical study.
It is also of interest to extend our formulation to include the
finite box size and investigate how the resonance effect is described in
Hill's system. 

\acknowledgments
The authors thank T. Takeuchi, E. Kokubo, S. Ida, and T. Tanigawa for
helpful discussions. 
%%%%%%%Added in Revision%%%%%%%%%
  We also thank the referee of this paper, Dr. Stuart Weidenschilling,
  for useful comments that improved this paper.  
%%%%%%%Added in Revision end%%%%%%%%%
This work was supported by the Grant-in-Aid for the Global COE Program
``The Next Generation of Physics, Spun from Universality and Emergence''
from the Ministry of Education, Culture, Sports, Science and Technology
(MEXT) of Japan.
The numerical calculations were carried out on Altix3700 BX2 at YITP in
Kyoto University. 
T. M. is supported by Grants-in-Aid for JSPS Fellows (19$\cdot$2409)
from MEXT of Japan.
S. I. is
supported by Grants-in-Aid (15740118, 16077202, and 18540238) from MEXT
of Japan.

\appendix

\section{Derivation of Equation \eqref{fy_integral}}
\label{App:fy_int}

In this section, we briefly show the derivation of \eqref{fy_integral}.
The method is the same as that of Goldreich and Tremaine (1980),
H\'{e}non and Petit (1986), or Hasegawa \& Nakazawa (1990).

We consider the restricted three-body problem.  In other words, we
consider the problem with $\nu=0$.  In this case, 
Jacobi energy $E_J$ conserves throughout the particle orbit.  Jacobi
energy is given by
\begin{equation}
 E_J = \dfrac{1}{2} r_{\rm p}^2 \Omega_{\rm p}^2 (h^2 + k^2) -
  \dfrac{3}{8} \Omega_{\rm p}^2 b^2 + \psi_{\rm p},
\end{equation}
where $\psi_{\rm p}$ is the gravitational potential of the planet.
Therefore, orbital elements before and after the encounter with the
planet are related by
\begin{equation}
 - \dfrac{3}{8} \Omega_{\rm p}^2 b^2 = \dfrac{1}{2} r_{\rm p}^2
  \Omega_{\rm p}^2 (h(\infty)^2 + k(\infty)^2) - \dfrac{3}{8}
  \Omega_{\rm p}^2 (b+\Delta b)^2.
\end{equation}
Assuming $\Delta b \ll b$, we obtain
\begin{equation}
 \Delta b \sim \dfrac{2 r_{\rm p}^2}{3b}
  \left( h(\infty)^2 + k(\infty)^2 \right).
\end{equation}
On the other hand, from equation \eqref{eq_b}, $\Delta b$ is given by
\begin{equation}
 \Delta b = \dfrac{2}{\Omega_{\rm p}} \int_{-\infty}^{\infty} F_y(t) dt.
  \label{db_fy}
\end{equation}
Therefore, we obtain
\begin{equation}
 \int_{-\infty}^{\infty} F_y(t) dt = \dfrac{r_{\rm p}^2 \Omega_{\rm
  p}}{3b} ( h(\infty)^2 + k(\infty)^2).
  \label{fyderive}
\end{equation}

From equation \eqref{sol_h} and \eqref{sol_k}, 
the elements $h$ and $k$ after the encounter are given by
\begin{equation}
 h(\infty) = - \dfrac{1}{r_{\rm p} \Omega_{\rm p}} \int_{-\infty}^{\infty} du 
  \left\{ F_x(u) \sin [ \Omega_{\rm p}u ] + 2F_y(u) 
   \cos [ \Omega_{\rm p} u ]  \right\}
  \label{app_h}
\end{equation}
\begin{equation}
 k(\infty) = \dfrac{1}{r_{\rm p} \Omega_{\rm p}} \int_{-\infty}^{\infty} du 
  \left\{ F_x(u) \cos [ \Omega_{\rm p}u ] - 2F_y(u) 
   \sin [ \Omega_{\rm p} u ]  \right\} .
  \label{app_k}
\end{equation}
Approximating the trajectory of the particle by circular orbit, we have
\begin{equation}
 F_x(t) \sim - \mathrm{sgn}(b) \dfrac{GM_{\rm p}}{b^2} 
  \dfrac{1}{\left( 1+(9/4)(\Omega_{\rm p}t)^2 \right)^{3/2}}
  \label{app_fx}
\end{equation}
\begin{equation}
 F_y(t) \sim \mathrm{sgn}(b) \dfrac{GM_{\rm p}}{b^2} 
  \dfrac{(3/2)\Omega_{\rm p}t}{\left( 1+(9/4)(\Omega_{\rm p}t)^2
			       \right)^{3/2}}.
  \label{app_fy}
\end{equation}
Integrating \eqref{app_h} and \eqref{app_k}, we have
\begin{equation}
 h(\infty) = 0
\end{equation}
\begin{equation}
 k(\infty) = - \mathrm{sgn}(b) \dfrac{8}{9} \dfrac{1}{r_{\rm
  p}\Omega_{\rm p}^2} \dfrac{GM_{\rm p}}{b^2} 
  \left[K_1\left(\dfrac{2}{3}\right) + 2 K_0 \left( \dfrac{2}{3} \right)
  \right].
\end{equation}
Substituting these results into equation \eqref{fyderive}, we finally
obtain
\begin{equation}
 \int_{-\infty}^{\infty} F_y(t) dt = \dfrac{64}{243} \dfrac{G^2 M_{\rm
  p}^2}{b^5 \Omega_{\rm p}^3} \left[ K_1\left( \dfrac{2}{3} \right) + 2
			      K_0 \left( \dfrac{2}{3} \right)\right]^2
\end{equation}
We note that direct substitution of equation \eqref{app_fy} into
equation \eqref{db_fy}  results in
zero, indicating that higher order of the expansion is essential.

\section{Proof of Equation \eqref{Dx_vanish}}
\label{app:pathline}

In this section, we show that the right hand side of equation
\eqref{Dx_calc} vanishes up to second order perturbation and prove
equation \eqref{Dx_vanish}.   The equation we prove is
\begin{eqnarray}
&  -\dfrac{2}{\Omega_{\rm p}} \displaystyle{\int_0^T} dt 
 \delta v_x^{(1)} \dfrac{\partial}{\partial x} \delta v_y^{(1)} 
  + \displaystyle{\int_0^T} dt_1 \dfrac{\partial 
  \delta v_x^{(1)}}{\partial x} (x_c,y_c(t_1))  
    \displaystyle{\int_0^{t_1}} dt_2 \delta v_x^{(1)} (x_c,y_c(t_2)) 
    \nonumber \\ 
&  + \displaystyle{\int_0^T} dt_1 \dfrac{\partial 
 \delta v_x^{(1)}}{\partial y} (x_c,y_c(t_1))  
    \displaystyle{\int_0^{t_1}} dt_2 \delta v_y^{(1)} (x_c,y_c(t_2)) = 0 
\label{Dx_calc_app}
\end{eqnarray}
We Fourier transform the perturbation in the $y$-direction,
\begin{equation}
 \delta f (x,y) = \overline{\delta f} + 
  \sum_{k_y \neq 0} \tilde{\delta f}_{k_y}(x) e^{ik_y y}
\end{equation}
where $\delta f$ denotes any perturbation quantity, 
$\overline{\delta f}$ is the average over $y$, and $\tilde{f}_{k_y}$ is
the Fourier component of non-axisymmetric modes.  Since physical
quantities are real, 
$\tilde{\delta f}_{k_y} = \tilde{\delta f}_{-k_y}^{\ast}$. 
In this section, we
drop superscript $(1)$ since all the perturbed values refer to first
order values.

Since we integrate over the circular orbit, integration with respect to
$t$ is converted to that with respect to $y$ by
\begin{equation}
 y = -\dfrac{3}{2} x_c \Omega_{\rm p} t.
\end{equation}
It is possible to show that the right hand side of equation
\eqref{Dx_calc_app} is equal to
\begin{eqnarray}
& -\dfrac{8L_y}{3\Omega_{\rm p}x_c} \sum_{k_y>0}
  \mathrm{Re} \left[ \tilde{d\delta v_x}_{k_y} \dfrac{\tilde{\delta 
	       v_y}_{k_y}^{\ast}}{dx} \right] 
 - \dfrac{8L_y}{9\Omega_{\rm p}^2 x_c^2} \sum_{k_y>0}
 \dfrac{1}{k_y} \mathrm{Im} \left[\dfrac{d \tilde{\delta
			     v_x}_{k_y}^{\ast}}{dx} \tilde{\delta
			     v_x}_{k_y} \right] \nonumber \\
& + \dfrac{8L_y}{9 \Omega_{\rm p}^2 x_c^2} \sum_{k_y>0}
 \left[ \tilde{\delta v_x}_{k_y}^{\ast} \tilde{\delta v_y}_{k_y}
 \right]
 + \dfrac{4L_y}{9\Omega_{\rm p}^2 x_c^2}
 \overline{\delta v_y} \delta v_x (x_c, L_y/2) .
\end{eqnarray}
The last term is zero as long as the box size is large enough.  Using
equation of continuity
\begin{equation}
 -\dfrac{3}{2}ik_y x \Omega_{\rm p} \dfrac{\tilde{\delta
  \Sigma}_{k_y}}{\Sigma_0} 
 + \dfrac{d}{dx}\tilde{\delta v_x}_{k_y} + ik_y \tilde{\delta v_y} =0 
\end{equation}
and conservation of vorticity
\begin{equation}
 \dfrac{d}{dx} \tilde{\delta v_y}_{k_y} - ik_y \tilde{\delta v_x}_{k_y}
  - \dfrac{1}{2} \Omega_{\rm p} \dfrac{\tilde{\delta
  \Sigma}_{k_y}}{\Sigma_0} = 0,
\end{equation}
we can show that the first three terms cancel.  Hence,
equation \eqref{Dx_vanish} is proved.

\section{Mass Flux of Sound Wave Propagating in a Homogeneous,
 Static Medium}
\label{app:soundwave}

In this section, we demonstrate that the sound wave that propagates in a
homogeneous, non-rotating medium carries mass flux, and therefore, show
that the vanishing mass flux given by equation \eqref{massflux_vanish}
is not a general conclusion but specific to the problem in
consideration.  We consider one-dimensional system which extends in
$x>0$, and at $x=0$, there is a forcing that creates a sound wave, which
is turned on at $t=0$.  The full system of equations are the equation of
continuity and Euler equation in one-dimensional system
\begin{eqnarray}
& \dfrac{\partial \rho}{\partial t} + \dfrac{\partial}{\partial x}(\rho
 x) = 0 \\
& \dfrac{\partial v}{\partial t} + v\dfrac{\partial v}{\partial x} =
 - \dfrac{c^2}{\rho} \dfrac{\partial \rho}{\partial x} + S\cos(\Omega t)
 \delta_D(x)\theta(t) ,
\end{eqnarray} 
where $\rho$ is density, $v$ is velocity, $c$ is sound speed, $S$ is the
amplitude of the source, $\Omega$ is the frequency of the forcing,
$\delta_D(x)$ is the Dirac's delta function, and $\theta(t)$ is step
function.  We consider background state with constant density $\rho_0$
and vanishing velocity.

The first order fluctuation caused by the forcing is given by
\begin{eqnarray}
& \dfrac{\partial}{\partial t} \dfrac{\delta \rho^{(1)}}{\rho_0} +
  \dfrac{\partial}{\partial x} \delta v^{(1)} = 0 \\
& \dfrac{\partial}{\partial t} \delta v^{(1)} + c^2
 \dfrac{\partial}{\partial x} \dfrac{\delta \rho}{\rho_0} = S
 \cos(\Omega t) \delta(x) \theta(t)
\end{eqnarray} 
and the second order perturbation is caused by the first order
perturbation as follows
\begin{eqnarray}
& \dfrac{\partial}{\partial t} \dfrac{\delta \rho^{(1)}}{\rho_0} +
 \dfrac{\partial}{\partial x} \delta v^{(1)} =  
 - \dfrac{\partial}{\partial x} \left[ \dfrac{\delta \rho^{(1)}}{\rho_0}
				\delta v^{(1)} \right] \\
& \dfrac{\partial}{\partial t} \delta v^{(1)} + c^2
 \dfrac{\partial}{\partial x} \dfrac{\delta \rho}{\rho_0} = 
 \dfrac{1}{2}\dfrac{\partial}{\partial x}
\left[  
 c^2 \left(\dfrac{\delta \rho^{(1)}}{\rho_0}\right)^2 
 - \delta v^{(1)^2}  \right].
\end{eqnarray}
Mass flux $F_M=\rho v$ is given by, up to the second order,
\begin{equation}
 F_M = \rho_0 \delta v^{(1)} + \rho_0 \delta v^{(2)} 
  + \delta \rho^{(1)} \delta v^{(1)}
\end{equation}

Since the sound wave propagates to $x\to\infty$ and there is no
reflection, the appropriate solution for the first order is given by
\begin{eqnarray}
& \dfrac{\delta \rho^{(1)}}{\rho_0} 
 = \dfrac{S}{c^2} \cos(\Omega t - kx),   \\
& \delta v^{(1)} = \dfrac{S}{c} \cos (\Omega t - kx)
\end{eqnarray}
where $k=\Omega/c$ and the boundary condition at $x=0$ is considered.

Equation for second order velocity fluctuation is given by
\begin{equation}
 \left[ -\dfrac{1}{c^2} \dfrac{\partial^2}{\partial t^2} +
			 \dfrac{\partial^2}{\partial x^2} \right] 
 \delta v^{(2)} = 
 2\dfrac{k^2 S^2}{c^3} \cos \left[ 2(\Omega t - kx) \right].
\end{equation}
Since all the perturbation must vanish at $t=0$, we obtain
\begin{equation}
 \delta v^{(2)} = - \dfrac{S^2}{2c^3} \Omega t 
  \sin \left[2(\Omega t - kx)  \right] .
\end{equation}
Therefore, mass flux is given by
\begin{equation}
 F_M = \dfrac{\rho_0 S^2}{2c^3} 
  + \dfrac{\rho_0 S}{c} \cos \left[ \Omega t - kx \right]
  + \dfrac{\rho_0 S^2}{2c^3} \Omega t 
    \sin \left[2(\Omega t -kx) \right]
  + \dfrac{\rho_0 S^2}{2c^3} \cos \left[ 2(\Omega t - kx) \right] .
\end{equation}
Therefore, there is a positive and finite mass
flux $\rho_0 S^2/2c^3$ remains even after spatial and temporal averages 
are taken.  It is also possible to show that
a fluid element move to $x>0$ on average by calculating path line.

\section{Interpolation Methods }
\label{app:sphint}

In the numerical calculation of large drag coefficient, stopping time of
the particle $\nu^{-1}$ is very small that the particle remains in the
same grid cell of the hydrodynamic calculation for several time steps.
In this situation, it is necessary to have a smooth data of gas velocity
even in sub-grid scale in order to obtain a smooth results.  In this
section, we describe how this can be realized.

Our method of interpolation is to consider the value of one
grid cell as a representative value of physical data around the grid
cell with a certain width, and add the contribution from nearby grid
cells to find the value of physical quantities at a location of
interest.  Let $f(x)$ be the value of a physical quantity of interest
and the data of $f(x)$ is given at discrete set of grid cells at
$x=x_i$.  We denote the data of $f(x)$ at $x=x_i$ by $f_i=f(x_i)$ and
$\Delta$ by grid size $\Delta=x_{i+1}-x_{i}$.  Let
the grid cell which is closest to the location of interest $x$ be
$x_{i0}$.  Our method of interpolation approximate the value of
$f$ at $x$ by
\begin{equation}
 f(x) \sim \dfrac{\sum_{i=i0-I}^{i=i0+I} f_i \exp[-(x_i/\eta
  \Delta)^2]}{\sum_{i=i0-I}^{i=i0+I}\exp[-(x_i/\eta \Delta)^2]}
\end{equation}
where $I$ is an integer and $\eta$ is a numerical factor.  If $\eta$ is
large, we have a more smoothed but more damped value.  We find
$I=20$ and $\eta=4$ gives a reasonably smooth results.

\clearpage

%%%%%%%%%%Figure changed in revision%%%%%%%%%%
%%%%%%%%%%Figure changed in revision end%%%%%%%%%%

%%%%%%%figure added in revision%%%%%%%%
%%%%%%%figure added in revision end%%%%%%%%

%%%%%%%figure added in revision%%%%%%%%
%%%%%%%figure added in revision end%%%%%%%%

%%%%%%%figures added in revision%%%%%%%
%%%%%%%figures added in revision end%%%%%%%

%%%%%%%figures added in revision%%%%%%%
%%%%%%%figures added in revision end%%%%%%%

%%%%%figures added in revision%%%%%%
%%%%%figures added in revision end%%%%%%

%%%%%figures added in revision%%%%%%
%%%%%figures added in revision end%%%%%%

%%%%%figures added in revision%%%%%%
%%%%%figures added in revision end%%%%%%

%%%%%figures added in revision%%%%%%
%%%%%figures added in revision end%%%%%%

\end{document}